\documentclass[a4paper,11pt]{article}
\usepackage[utf8]{inputenc}
\usepackage{graphicx}
\usepackage{hyperref}
\usepackage{latexsym}
\usepackage{color}
\usepackage{amsmath}
\usepackage{geometry}
\usepackage{slashed}

\definecolor{green}{rgb}{0,.5,0}

\definecolor{red}{rgb}{1,0,0}

\def\Tr{{\rm Tr}}
\def\msbar{{\overline{\rm MS}}}

\usepackage{multirow}

\title{RI/(S)MOM renormalizations of overlap quark bilinears with different levels of hypercubic smearing}
\author{Yujiang Bi$^1$, Ying Chen$^{1,2}$, Ming Gong$^{1,2}$, Fangcheng He$^{3,4}$, Keh-Fei Liu$^{5}$,\\
Zhaofeng Liu$^{1,2,6}$\thanks{liuzf@ihep.ac.cn}, Yi-Bo Yang$^{2,3,7,8}$, Dian-Jun Zhao$^{2,3}$\\
\vspace*{0.4cm}
\includegraphics[scale=0.20]{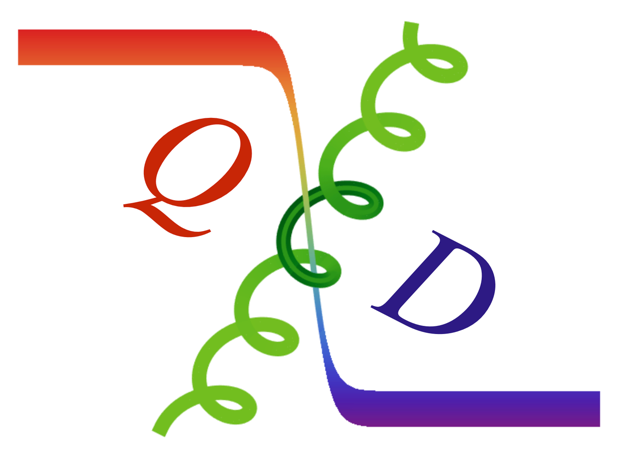}\\
\vspace*{0.4cm}
 ($\chi$QCD Collaboration)
}
\date{}

\begin{document}

\maketitle
\begin{center}
$^1$Institute of High Energy Physics, Chinese Academy of Sciences, Beijing 100049, China\\
$^2$School of Physics, University of Chinese Academy of Sciences, Beijing 100049, China\\
$^3$CAS Key Laboratory of Theoretical Physics, Institute of Theoretical Physics, Chinese Academy of Sciences, Beijing 100190, China\\
$^4$Department of Physics and Astronomy, Stony Brook University, Stony Brook, New York 11794, USA\\
$^5$Department of Physics and Astronomy, University of Kentucky, Lexington, Kentucky 40506, USA\\
$^6$Center for High Energy Physics, Peking University, Beijing 100871, China\\
$^{7}$School of Fundamental Physics and Mathematical Sciences, Hangzhou Institute for Advanced Study, UCAS, Hangzhou 310024, China\\
$^{8}$International Centre for Theoretical Physics Asia-Pacific, Beijing/Hangzhou, China\\
\end{center}

\begin{abstract}
On configurations with 2+1-flavor dynamical domain-wall fermions, we calculate the RI/(S)MOM
renormalization constants (RC) of overlap quark bilinears.
Hypercubic (HYP) smearing is used to construct the overlap Dirac operator. We investigate the possible effects of the smearing on discretization errors in the RCs by varying the level of smearing
from 0 to 1 and 2. The lattice is of size $32^3\times64$ and with lattice spacing $1/a=2.383(9)$ GeV. The RCs in the $\msbar$ scheme at 2 GeV are given at the end, with
the uncertainty of $Z_T$ reaching $\le1$\% for the tensor current. Results of the renormalized quark masses and hadron matrix elements show that the renormalization procedure suppresses the $\sim$ 30\% difference of the bare quantities with or without HYP smearing into the 3\%-5\% level.
\end{abstract}

\section{Introduction}
The uncertainties of renormalization constants (RCs) can be a main source of the uncertainties of
hadronic matrix elements as lattice quantum chromodynamics (LQCD) is used to precisely calculate
matrix elements numerically. For example, in our previous work on meson decay constants~\cite{Chen:2020qma} the uncertainties of
the tensor decay constants of $D^*_{(s)}$ are dominated by the uncertainty of the RC of the tensor current ($Z_T$). Therefore it is crucial to compute RCs as precise as possible.
In Ref.~\cite{Hatton:2020vzp} the authors used momentum-subtraction schemes as intermediate schemes to determine $Z_T$ very precisely. After removing the nonperturbative effects from condensate contaminations they
managed to obtain a precision better than 0.5\% for $Z_T$ in the $\msbar$ scheme at the scale
of the bottom quark pole mass.

The momentum-subtraction scheme SMOM~\cite{Aoki:2007xm,Sturm:2009kb} was introduced to reduce the uncertainties of the RCs of quark bilinear operators, especially for the scalar density ($Z_S$)
and therefore for the quark mass ($Z_m$) through $Z_m=1/Z_S$ for lattice chiral fermions. This scheme uses a symmetric combination for the momenta on the legs of the Green functions of the operators under renormalization. It is expected to have less nonperturbative infrared contaminations in the pseudoscalar and axial vector vertex functions
compared with the MOM scheme~\cite{Martinelli:1994ty} which uses forward Green functions.
The perturbative conversion ratio of $Z_S$ from the SMOM scheme to the $\msbar$ scheme was shown~\cite{Gorbahn:2010bf,Almeida:2010ns,Kniehl:2020sgo,Bednyakov:2020ugu} to converge faster than that for the MOM scheme. Thus the truncation error from the perturbative conversion is reduced significantly.

In our previous work~\cite{Bi:2017ybi} on renormalization of quark bilinears for overlap fermions on domain-wall fermion configurations we compared SMOM and MOM schemes numerically. We indeed see that the infrared effects in the pseudoscalar and axial vector vertex functions are more suppressed in the SMOM scheme and $Z_S=Z_P$ and $Z_V=Z_A$ can be better satisfied. However we had difficulties to describe the $a^2p^2$ dependence of $Z_{S,P}^{\msbar}(2\mbox{ GeV}; a^2p^2)$ obtained through the SMOM scheme, where $a^2p^2$ is the
initial scale in lattice units at which we do the renormalization and from which we run $Z_{S,P}$ to 2 GeV.
It seems that we could not find a good renormalization window in $a^2p^2$ in which both the nonperturbative effects and lattice discretization effects are small. While if we use the MOM scheme, a good window for $Z_{S,P}^{\msbar}(2\mbox{ GeV}; a^2p^2)$ can be found,
in which the $a^2p^2$ dependence can be well fitted by a linear function and is attributed to
$\mathcal{O}(a^2p^2)$ discretization effects. Similar observations were made in a more recent work~\cite{He:2022lse}, in which both the SMOM and MOM schemes were used to obtain the RCs for
overlap quark bilinears on both domain-wall fermion and highly improved staggered quark configurations.

In the mixed action setup of the $\chi$QCD collaboration with overlap fermions on domain-wall fermion configurations, one level of hypercubic (HYP) smearing~\cite{Hasenfratz:2001hp} on the gauge links is performed to construct the overlap Dirac operator. This can expedite the inversions of the Dirac operator~\cite{xQCD:2010pnl}. Link smearing is widely used in LQCD simulations
for various lattice fermions and has many benefits. For example, HYP smearing can suppress chiral symmetry breaking effects in Wilson fermions besides speeding up inversions for
overlap fermions~\cite{Hasenfratz:2007rf}.
In Ref.~\cite{Arthur:2013bqa} the authors studied the effects of smearing on vertex functions when momentum-subtraction schemes are used for renormalization. The upper end of the
renormalization window is observed to be lowered by link smearing.

In this work we try to investigate the relation between the HYP smearing that we use to construct our overlap Dirac operator and the $a^2p^2$ dependence of $Z_{S}^{\msbar}(2\mbox{ GeV}; a^2p^2)$ and also of other RCs. Both the MOM and SMOM schemes are used as intermediate schemes and the results are compared. The configurations used here were generated by the
RBC/UKQCD collaborations with volume $32^3\times64$ and lattice spacing $1/a=2.383(9)$ GeV (labeled as 32I)~\cite{RBC:2010qam}. This ensemble was also used in the study of~\cite{Arthur:2013bqa}.
We calculate the RCs for the tensor current and the quark field ($Z_q$) in the $\msbar$ scheme on the 32I ensemble
that were not done in our previous work~\cite{Liu:2013yxz}. The new three-loop conversion ratio for $Z_T$ from the SMOM
scheme to the $\msbar$ scheme~\cite{Kniehl:2020sgo} and four-loop ratios for $Z_{S,T,q}$
from MOM to $\msbar$ are used in this work. Thus the systematic uncertainties of $Z_{S,T}$
 are now reduced compared with those in Refs.~\cite{Bi:2017ybi,Liu:2013yxz}.

The paper is organized as follows. In Sec.~\ref{sec:setup} we describe our lattice setup
and briefly refresh the formulas of the MOM and SMOM schemes.
Then in Sec.~\ref{sec:results} we give the details of our calculation, the comparison of
the results from different levels of HYP smearing and different intermediate schemes, and
the discussions. We also investigate the renormalizated meson two-point functions at several quark masses in Sec.~\ref{sec:twopt}, to verify our renormalization procedure and estimate the discretization error due to different HYP smearing steps. Finally we summarize in Sec.~\ref{sec:summary}

\section{Methodology}
\label{sec:setup}
\subsection{Simulation parameters}
We use the 2+1-flavor gauge configurations generated by the RBC/UKQCD collaborations~\cite{RBC:2010qam}.
The sea quarks are domain-wall fermions with masses $am_l=0.004$, $0.006$, $0.008$ and $am_s=0.03$ for the mass degenerate
up and down quarks and the strange quark, respectively. The parameters of the configurations
are listed in Table~\ref{tab:nconfs}. More detailed information can be found in Ref.~\cite{RBC:2010qam}.
\begin{table}
\begin{center}
\caption{Parameters of configurations used in this work. The residual mass of the domain-wall fermion in lattice units $am_{\rm res}$
is in the two-flavor chiral limit as given in Ref.~\cite{Aoki:2010dy}.}
\begin{tabular}{ccccc}
\hline\hline
Label & $am_l/am_s$ & Volume & $N_{\rm conf}$ & $am_{\rm res}$ \\
\hline
f004 & 0.004/0.03 & $32^3\times64$ & 42 & 0.0006664(76) \\
f006 & 0.006/0.03 & $32^3\times64$ & 42 & \\
f008 & 0.008/0.03 & $32^3\times64$ & 49 & \\
\hline\hline
\end{tabular}
\label{tab:nconfs}
\end{center}
\end{table}

For the valence quark we use overlap fermions. The massless overlap Dirac operator~\cite{Neuberger:1997fp} is defined as
\begin{equation}
D_{\rm ov}  (\rho) =   1 + \gamma_5 \varepsilon (\gamma_5 D_{\rm w}(\rho)),
\label{eq:Dov}
\end{equation}
where $\varepsilon$ is the matrix sign function and $D_{\rm w}(\rho)$ is the usual Wilson fermion operator,
except with a negative mass parameter $- \rho = 1/2\kappa -4$ in which $\kappa_c < \kappa < 0.25$.
We use $\kappa = 0.2$ in our calculation which corresponds to $\rho = 1.5$. The massive overlap Dirac operator is defined as
\begin{eqnarray}
D_m &=& \rho D_{\rm ov} (\rho) + m\, (1 - \frac{D_{\rm ov} (\rho)}{2}) \nonumber\\
       &=& \rho + \frac{m}{2} + (\rho - \frac{m}{2})\, \gamma_5\, \varepsilon (\gamma_5 D_{\rm w}(\rho)).
\end{eqnarray}
To accommodate the SU(3) chiral transformation, it is usually convenient to use the chirally regulated field
$\hat{\psi} = (1 - \frac{1}{2} D_{\rm ov}) \psi$ in lieu of $\psi$ in the interpolation field and the currents.
This is equivalent to leaving the unmodified currents and instead adopting the effective propagator,
\begin{equation}
G \equiv D_{\rm eff}^{-1} \equiv (1 - \frac{D_{\rm ov}}{2}) D^{-1}_m = \frac{1}{D_c + m},
\end{equation}
where $D_c = \frac{\rho D_{\rm ov}}{1 - D_{\rm ov}/2}$ satisfies $\{\gamma_5, D_c\}=0$~\cite{Chiu:1998gp}.
With the good chiral properties of overlap fermions,
we can expect $Z_S=Z_P$ and $Z_V=Z_A$. These relations were verified in our previous works on renormalizations of those bilinear quark operators~\cite{Bi:2017ybi,Liu:2013yxz}.

We apply no smearing, one and two levels of HYP smearing in constructing the overlap Dirac operator as given in Eq.(\ref{eq:Dov}). In the rest of this paper the three cases are labeled as ``thin," ``HYP1," and ``HYP2" respectively. Bootstrap and jackknife analyses are used to
obtain the statistical uncertainties.

Same ten valence quark masses in lattice units are used for all smearing cases and ensembles. Their values are given in Table~\ref{tab:amqv}. The corresponding pion masses on the ensemble f004
are also given in Table~\ref{tab:amqv}, which are in the range of 220 to 500 MeV and are obtained by fitting two-point functions as described in Sec.~\ref{sec:zawi}. We use more statistics for the HYP1 case which corresponds to the setup used in most of the $\chi$QCD studies, and then the uncertainty of $am_\pi$ for this case is smaller. For the HYP1 case, we also list the $m_{\pi}L$ in the Table. The fitted pion masses at a same bare quark mass differ by renormalization effects and discretization errors.

\begin{table}
\begin{center}
\caption{Valence quark masses in lattice units. The corresponding pion masses are from fittings
to two-point functions on ensemble f004
as explained in Sec.~\ref{sec:zawi}. Only statistical errors are given.}
\begin{tabular}{cc|ccccc}
\hline\hline
$am_q$ &  & 0.00460 & 0.00585 & 0.00677 & 0.00765 & 0.00885 \\
\multirow{3}{*}{$am_\pi$} & Thin  & 0.0850(17) & 0.0954(17) & 0.1027(21) & 0.1090(17) & 0.1172(16) \\
& HYP1 & 0.0955(07) & 0.1073(07) & 0.1152(07) & 0.1222(07) & 0.1310(07) \\
& HYP2 & 0.0895(22) & 0.1026(21) & 0.1108(20) & 0.1173(20) & 0.1272(19) \\
$m_\pi L$ & HYP1 & 3.1 & 3.4 & 3.7 & 3.9 & 4.2 \\
\hline
$am_q$ &  & 0.01120   & 0.01290 & 0.01520 & 0.01800 & 0.02400 \\
\multirow{3}{*}{$am_\pi$} & Thin & 0.1318(19) & 0.1413(19) & 0.1531(19) & 0.1664(18) & 0.1889(18) \\
& HYP1  & 0.1466(06) & 0.1569(06) & 0.1694(06) & 0.1836(06) & 0.2115(05) \\
& HYP2 & 0.1432(19) & 0.1537(18) & 0.1666(18) & 0.1812(18) & 0.2091(18)\\ $m_\pi L$ & HYP1 & 4.7 & 5.0 & 5.4 & 5.9 & 6.8
\\
\hline\hline
\end{tabular}
\label{tab:amqv}
\end{center}
\end{table}

\subsection{Momentum subtraction schemes}
The MOM scheme~\cite{Martinelli:1994ty} is defined in the quark massless limit by imposing conditions on forward vertex functions $\Lambda_{\mathcal{O}}$ of quark bilinear
operators $\mathcal{O}$ at a renormalization scale $\mu$
\begin{equation}
\underset{m_q\to 0}{\text{lim}}
Z^{-1}_qZ_{\mathcal{O}}\frac{1}{12}\Tr[\Lambda_{\mathcal{O},B}(p)\Lambda_{\mathcal{O}}^{\text{tree}}(p)^{-1}]_{p^2=\mu^2}=1,
\label{eq:condition}
\end{equation}
where the subscript $B$ stands for bare and $\Lambda_{\mathcal{O}}$ can be calculated from the quark propagator $S(p)$ and the
Green function $G_{\mathcal{O}}(p_1,p_2)$ as
\begin{equation}
\Lambda_{\mathcal{O}}(p_1,p_2)=S^{-1}(p_1)G_{\mathcal{O}}(p_1,p_2)S^{-1}(p_2)
\label{eq:vertex}
\end{equation}
with $p_1=p_2=p$.
The projector $\Lambda_{\mathcal{O}}^{\rm tree}(p)=\Gamma$ for the quark bilinear operators
$\bar\psi\Gamma\psi$ ($\Gamma=I, \gamma_\mu\gamma_5, \sigma_{\mu\nu}(=\frac{1}{2}[\gamma_\mu,\gamma_\nu])$) considered in this work.
The trace ``Tr" is over the color and spin indices.
The quark field renormalization $Z_q$ is given by
\begin{equation}
Z_q^{\rm MOM}(\mu)=\underset{m_q\to 0}{\text{lim}}\frac{-i}{48}\Tr\left[\gamma_\nu\frac{\partial S^{-1}(p)}{\partial p_\nu}\right]_{p^2=\mu^2},
\end{equation}
so that the chiral Ward identities are satisfied in the MOM scheme and we have $Z_{A}^{\rm MOM}=Z_{A}^{\msbar}$.

For the SMOM scheme~\cite{Sturm:2009kb} one considers symmetric momentum combinations
\begin{equation}
q^2\equiv(p_1-p_2)^2=p_1^2=p_2^2=\mu^2
\label{eq:sym_mom}
\end{equation}
in Eq.(\ref{eq:vertex}). The renormalization conditions for the scalar and tensor currents
in the SMOM scheme are the same as Eq.(\ref{eq:condition}) but at the symmetric point Eq.(\ref{eq:sym_mom}). The conditions for the quark field and the axial current are now
\begin{equation}
  Z_q^{\rm RI/SMOM}=\underset{m_q\to 0}{\text{lim}}\frac{1}{12p^2}\text{Tr}[S_B^{-1}(p)p\!\!\!/]_{p^2=\mu^2}
  \label{eq:zq_smom}
\end{equation}
and
\begin{equation}
\underset{m_q\to 0}{\text{lim}}Z_q^{-1}Z_A\frac{1}{12q^2}\text{Tr}[q_{\mu}\Lambda_{A,B}^{\mu}(p_1,p_2)\gamma_5q\!\!\!/]_{\rm sym}=1,
\label{eq:za_smom}
\end{equation}
respectively.
Here the subscript ``sym" denotes the symmetric condition in Eq.(\ref{eq:sym_mom}).
The conditions in Eqs.(\ref{eq:zq_smom},\ref{eq:za_smom}) are also consistent with the
chiral Ward identities. Therefore one has $Z_{A}^{\rm SMOM}=Z_{A}^{\msbar}$.

The Green function $G_{\mathcal{O}}$ is computed between two external off-shell quark states.
Thus a gauge fixing (usually Landau gauge) is used to implement the renormalization conditions in both the MOM
and SMOM schemes. In this study we use
point source to calculate the quark propagator
\begin{equation}
  S(p)=\sum_{x} e^{-ip\cdot x}\langle \psi(x)\bar{\psi}(0)\rangle,
\end{equation}
and then
\begin{align}
  \begin{split}
	G_{\mathcal{O}}(p_1,p_2)=&\sum_{x,y}e^{-i(p_1\cdot x-p_2\cdot y)}\left<\psi(x)\mathcal{O}{(0)\bar{\psi}(y)}\right>.
  \end{split}
\end{align}

We use periodic boundary conditions in all four directions on our lattice when inverting the overlap Dirac operator. Thus the momentum modes in lattice units can be written as
\begin{equation}
ap=2\pi\left(\frac{k_1}{L}, \frac{k_2}{L}, \frac{k_3}{L}, \frac{k_4}{T}\right),
\end{equation}
where $L=32$, $T=64$ and $k_\mu$ are integers: $k_{1,2,3}=-8, -7, ..., 8$, $k_4=-16, -15, ..., 16$.

For the MOM scheme we apply a democratic cut on the momentum modes
\begin{equation}
\frac{p^{[4]}}{(p^2)^2}<0.26,\quad\mbox{where } p^{[4]}=\sum_\mu p_\mu^4,\quad p^2=\sum_\mu p_\mu^2
\label{eq:p_cut}
\end{equation}
to reduce the discretization effects from O(4) to H(4) symmetry breaking on the lattice.
Note this cut is more strict than those used in Refs.~\cite{Bi:2017ybi,Liu:2013yxz}.

It is difficult to apply the above democratic cut on all the symmetric momentum modes ($q^2=p_1^2=p_2^2$) required by the SMOM scheme.
But still we can apply a cut on both $p_1$ and $p_2$
\begin{equation}
\frac{p^{[4]}}{(p^2)^2}<0.34,\quad\mbox{for } p_1\mbox{ and } p_2.
\label{eq:smom_cut}
\end{equation}
That is to say $p_1$ and $p_2$ are almost body diagonal and $q$ is almost along an axis (to satisfy $q^2=p_1^2$). A typical example (with $p^{[4]}/(p^2)^2=0.25$) is
\begin{equation}
ap_1=\frac{2\pi}{L}(k,k,k,k),\quad ap_2=\frac{2\pi}{L}(-k,k,k,k),\quad aq=\frac{2\pi}{L}(2k,0,0,0).
\end{equation}
The cut 0.34 in Eq.(\ref{eq:smom_cut}) is looser than the 0.26 in Eq.(\ref{eq:p_cut}) because the symmetric condition $q^2=p_1^2$ should also be satisfied. If 0.26 is applied, too few momentum modes will be left. In this way the numerical results of RCs from the SMOM scheme will be less scattered around a smooth curve.

\section{Numerical results and discussions}
\label{sec:results}
\subsection{Local axial vector current}
\label{sec:zawi}
We start with the renormalization of the local axial vector current. The RC $Z_A$ is then
used to scale the other RCs for the quark field, the scalar density and the tensor current.
$Z_A$ can be obtained by using the partially conserved axial current relation
\begin{equation}
Z_A\partial_\mu A_\mu=2Z_m m_q Z_P P,
\label{eq:ZA_WI}
\end{equation}
where $A_\mu$ and $P$ are the axial vector current and pseudoscalar density, respectively,
and the quark mass RC $Z_m$ equals $Z_P^{-1}$ for lattice chiral fermions such as the overlap fermion used in this work.

Consider the zero momentum two-point correlators in the pseudoscalar channel
\begin{eqnarray}
C_{PP}(t)&\equiv &\sum_{\vec x}\langle\Omega|P(x)P^\dagger(0)|\Omega\rangle,\label{eq:cpp}\\
C_{A_4P}(t)&\equiv &\sum_{\vec x}\langle\Omega|A_4(x)P^\dagger(0)|\Omega\rangle.\label{eq:ca4p}
\end{eqnarray}
From Eq.(\ref{eq:ZA_WI}) one finds
\begin{equation}
Z_A^{\rm WI}=\frac{2m_q\langle\Omega|P|\pi\rangle}{m_\pi\langle \Omega| A_4 |\pi\rangle},
\label{eq:za_me}
\end{equation}
and
\begin{equation}
\sum_{\vec x}\langle \Omega|Z_A\partial_\mu A_\mu(x)P^\dagger(0)|\Omega\rangle=2 m_q \sum_{\vec x}\langle\Omega|P(x)P^\dagger(0)|\Omega\rangle,
\label{eq:pcac_corr}
\end{equation}
or
\begin{equation}
Z_A^{\rm WI}=\frac{4m_q C_{PP}(t)}{C_{A_4P}(t+1)-C_{A_4P}(t-1)},
\label{eq:corr}
\end{equation}
where the partial derivative is replaced by a difference.

One can simultaneously fit the two-point correlation functions in Eqs.(\ref{eq:cpp},\ref{eq:ca4p}) at large time, where the ground state
contribution dominates, and use Eq.(\ref{eq:za_me}) to calculate $Z_A^{\rm WI}$. Alternatively one
may use the ratio Eq.(\ref{eq:corr}) to get $Z_A^{\rm WI}$. We compared the results from the two methods in our previous work~\cite{Bi:2017ybi} and found that $Z_A$ from the first method
has less dependence on the valence quark mass. Thus, we choose to use the first method
in this work.

Two-point correlation functions on 42 configurations are calculated with a Z3 noise wall source at a given time slice for ensemble f004 for the three smearing cases.
Note no gauge fixing is needed for calculating the gauge invariant correlation functions.

\begin{figure}[htbp]
\centering \includegraphics[width=0.49\textwidth]{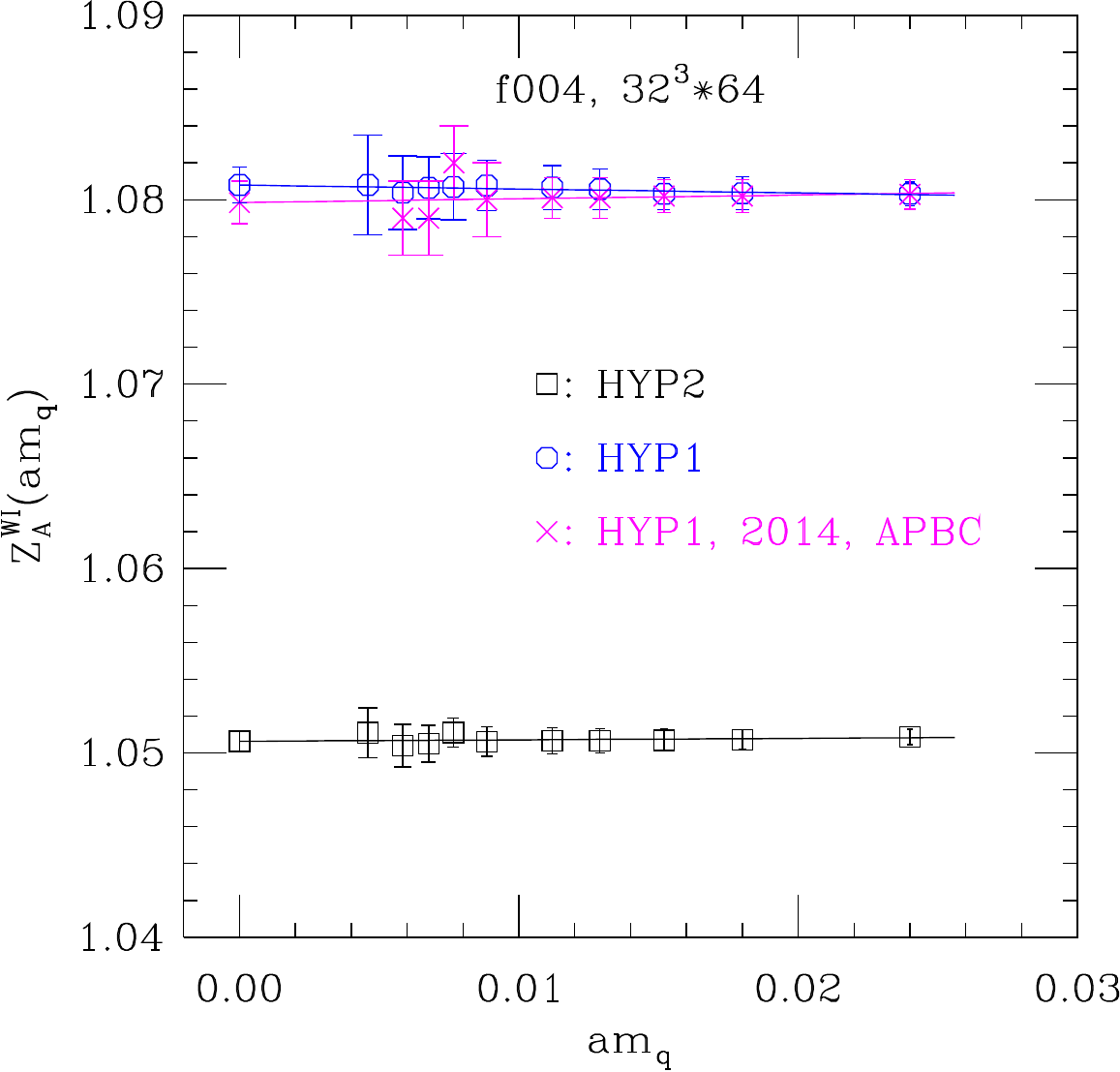}
\includegraphics[width=0.49\textwidth]{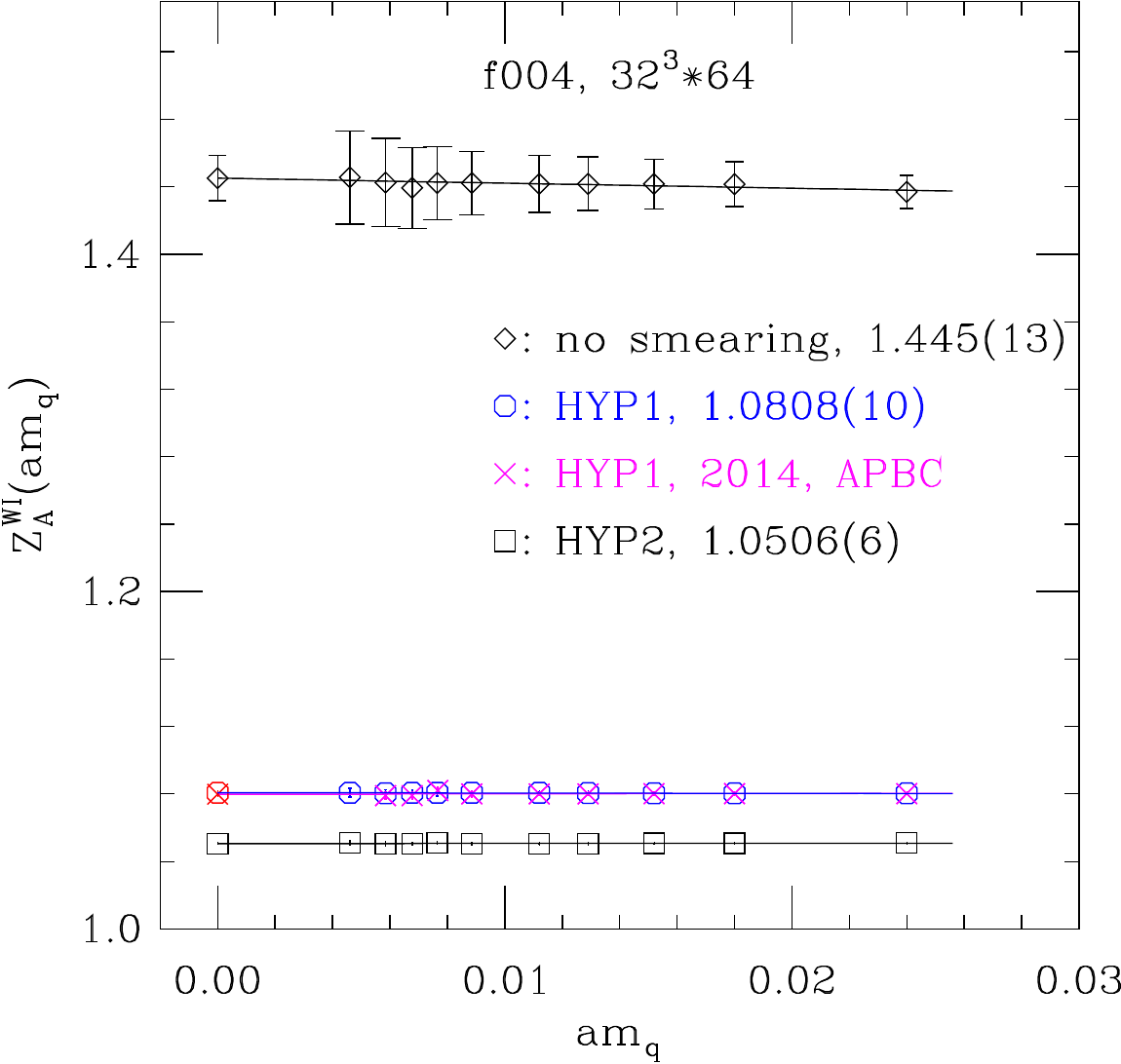}
 \caption{Left panel: $Z_A$ from Eq.(\ref{eq:za_me}) for HYP1 and HYP2. Right panel:
 $Z_A$ from Eq.(\ref{eq:za_me}) for all three smearing cases. The straight lines in both graphs are  linear chiral extrapolations of $Z_A$. The values in the chiral limit are given in the right graph. For the case HYP1 comparison is also shown with our previous work~\cite{Liu:2013yxz} with an antiperiodic boundary condition in the time direction.}
 \label{fig:za_corr}
\end{figure}

The results of $Z_A^{\rm WI}$ are collected in table~\ref{tab:zawi} and drawn in figure~\ref{fig:za_corr}.
\begin{table}[htbp]
\begin{center}
\caption{$Z_A^{\text{WI}}$ at ten valence quark masses for the three smearing cases on ensemble f004.
42 configurations are used to calculate two-point functions with a Z3 noise wall source.
The uncertainties are statistical and are from bootstrap analyses.}
\begin{tabular}{ccccccc}
\hline\hline
$am_q$ & & 0.00460 & \hspace{0.3em}0.00585 & \hspace{0.3em}  0.00677 &  \hspace{0.3em} 0.00765 &  \hspace{0.3em}0.00885  \\
 & Thin & 1.446(28) & 1.443(26) & 1.439(24) & 1.442(22) & 1.442(19) \\
$Z_A^{\rm WI}$ & HYP1 & 1.0808(27)    & 1.0804(20) & 1.0806(17) & 1.0807(18) & 1.0808(14)  \\
 & HYP2 & 1.0511(14) & 1.0504(12) & 1.0505(10) & 1.0511(8) & 1.0506(8) \\
\hline
  $am_q$ & &  \hspace{0.3em}0.01120 &  \hspace{0.3em}0.01290 & \hspace{0.3em} 0.01520 & \hspace{0.3em} 0.01800 & \hspace{0.3em} 0.02400  \\
  & Thin & 1.442(17) & 1.442(16) & 1.442(15) & 1.442(13) & 1.437(10) \\
$Z_A^{\rm WI}$ & HYP1 &  1.0807(12) & 1.0806(11) & 1.0803(9) & 1.0804(9) & 1.0803(6)\\
 & HYP2 & 1.0506(7) & 1.0507(7) & 1.0507(6) & 1.0507(5) & 1.0509(4) \\
\hline\hline
\end{tabular}
\label{tab:zawi}
\end{center}
\end{table}
To fairly compare the three smearing cases we use the same statistics in figure~\ref{fig:za_corr}.
We can clearly see that smearing decreases the statistical error of $Z_A^{\rm WI}$ and drives it
closer to one as the level of smearing increases. By linearly extrapolating $Z_A^{\rm WI}$
to the valence quark massless limit, one obtains
\begin{equation}
Z_A^{\rm WI}=1.445(13),\quad 1.0808(10),\quad 1.0506(6),
\label{eq:za_numbers}
\end{equation}
for thin, HYP1 and HYP2, respectively. Since the uncertainty of $Z_A^{\rm WI}$ using the thin link is much larger than the other cases, we repeat the calculation with a Coulomb wall source~\cite{He:2022lse} to improve the statistics, and obtain $Z_{A,{\rm thin}}^{\rm WI}=1.4403(6)$ which is consistent with that listed in Eq.~(\ref{eq:za_numbers}) but has much smaller uncertainty.

We also compare $Z_A^{\rm WI}$ for the case HYP1 obtained here with that from our previous work in figure~\ref{fig:za_corr}. One can see
that the results are in good consistency although different boundary conditions in the time direction
are used here and in~\cite{Liu:2013yxz}. This is not surprising since RCs are not sensitive to finite volume effects from different boundary conditions.

\begin{figure}[htbp]
\centering \includegraphics[width=0.49\textwidth]{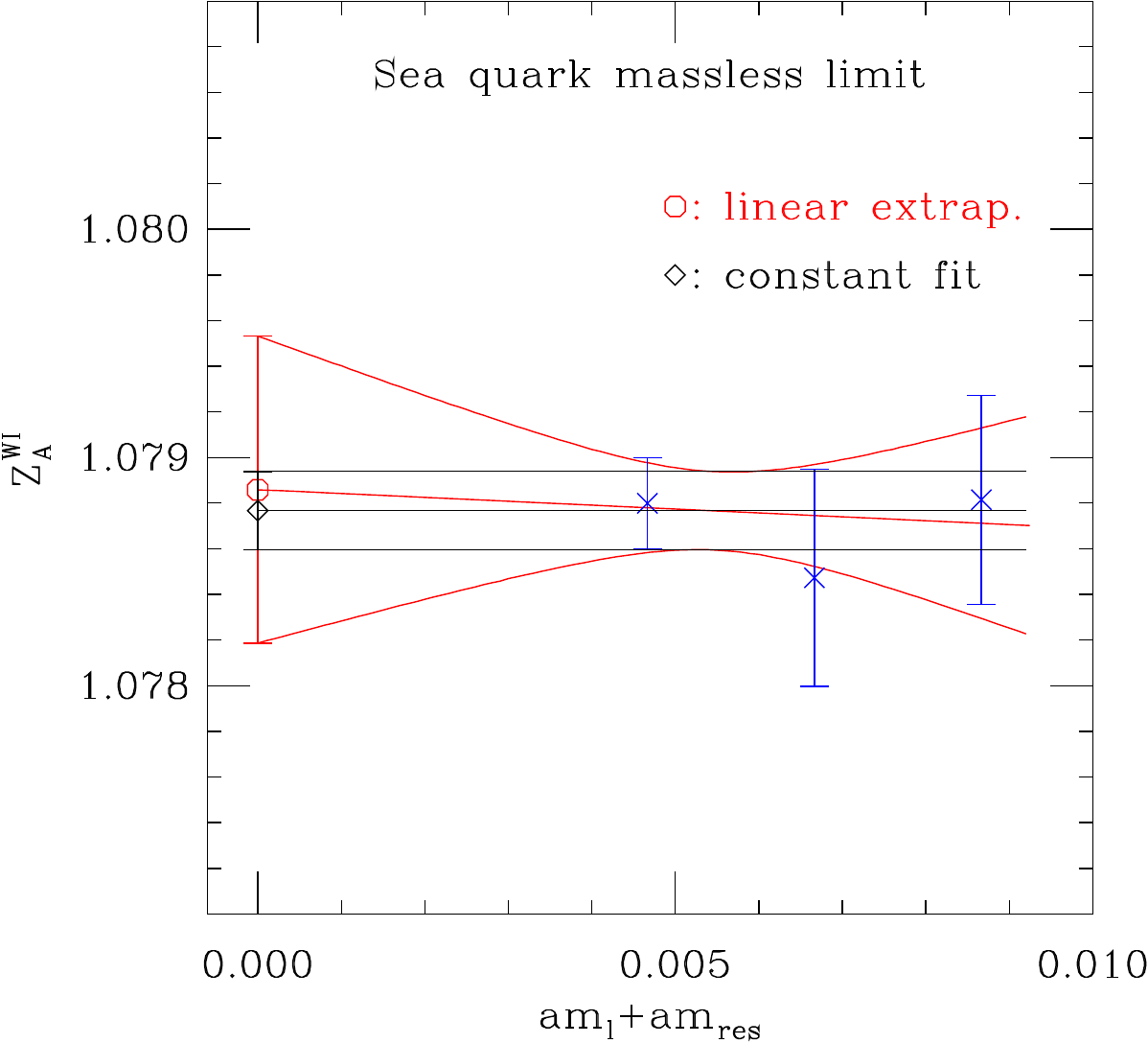}
 \caption{The linear extrapolation of $Z_A^{\rm WI}$ to the sea quark massless limit (red) and the constant fit (black).}
 \label{fig:za-sea}
\end{figure}

We will need the RCs for the HYP1 case for calculating hadronic matrix elements in the future.
To shrink the statistical errors for this smearing case we use 628 configurations, 16 sources on 42 and 16 sources on 49 configurations to compute the two-point correlators in Eqs.(\ref{eq:cpp},\ref{eq:ca4p}) on ensembles f004, f006 and f008, respectively. The linear chiral extrapolation of $Z_A^{\rm WI}(am_q)$ is similar to those shown in figure~\ref{fig:za_corr}. In the chiral limit of valence quark
we find
\begin{equation}
Z_A^{\rm WI}=1.0788(2),\quad 1.0785(5) \mbox{ and } 1.0788(5),
\end{equation}
 respectively, on the three ensembles. Here the error is statistical and we used all ten valence quark masses for doing the chiral extrapolations. For the chiral limit of the up and down quarks in the sea
we linearly extrapolate the results on the three ensembles to $am_l+am_{\rm res}=0$ and
obtain $Z_A=1.0789(7)$. This extrapolation as well as a constant fit (1.0788(2)) is shown in figure~\ref{fig:za-sea}. The difference between the linear extrapolation and the constant fit is taken as a systematic error (0.0001) below.
We repeat the above chiral extrapolations for the other three choices of the valence quark masses:
(1) remove the lightest four; (2) remove the heaviest two; (3) remove the lightest four and
the heaviest two. Then the largest variation in the center values is taken as a systematic uncertainty (0.0006).

Another systematic uncertainty comes from the massive strange quark in the sea. By using half
of the slope from the linear chiral extrapolation of the up/down sea quarks, we estimate the
change of $Z_A$ to be 0.0003 in the limit $am_s+am_{\rm res}=0$. In the end we have $Z_A=1.0789(7)(1)(6)(3)$. Adding up all the errors quadratically, one gets
$Z_A=1.0789(10)$.
This value agrees with our previous result 1.086(2)~\cite{Liu:2013yxz} at $\sim3\sigma$. We now have more statistics and consider more carefully the systematical uncertainties in our new result.

From the fitting of the two-point functions we also obtain the pion mass at each valence quark
mass as given in table~\ref{tab:amqv}.

Since the renormalization conditions in both the MOM and SMOM schemes are consistent with the
chiral Ward identities, we have $Z_A^{\rm MOM}=Z_A^{\rm SMOM}=Z_A^{\rm WI}$ in the continuum limit. On the lattice they can differ by discretization effects.

\subsection{Scalar density}
The scalar density renormalization constant $Z_S^{\rm MOM}$ can be obtained from
$Z_A^{\rm WI}$ computed in Sec.~\ref{sec:zawi} and the ratio of projected vertex functions,
\begin{equation}
\frac{Z_S^{\rm MOM}}{Z_A^{\rm MOM}}=\left.\frac{\Gamma_A(p)}{\Gamma_S(p)}\right|_{p^2=\mu^2},
\label{eq:zsza}
\end{equation}
where
\begin{equation}
\Gamma_A(p)=\frac{1}{48}\Tr[\Lambda_{A,B}^{\mu}(p)\gamma_5\gamma_\mu],
\quad \Gamma_S(p)=\frac{1}{12}\Tr[\Lambda_{S,B}^{\mu}(p)].
\end{equation}
The scale dependence of the ratio in Eq.(\ref{eq:zsza}) is governed by the anomalous dimension
of the scalar density since $Z_A$ is scale independent.

\begin{figure}[htbp]
\centering
\includegraphics[width=0.49\textwidth]{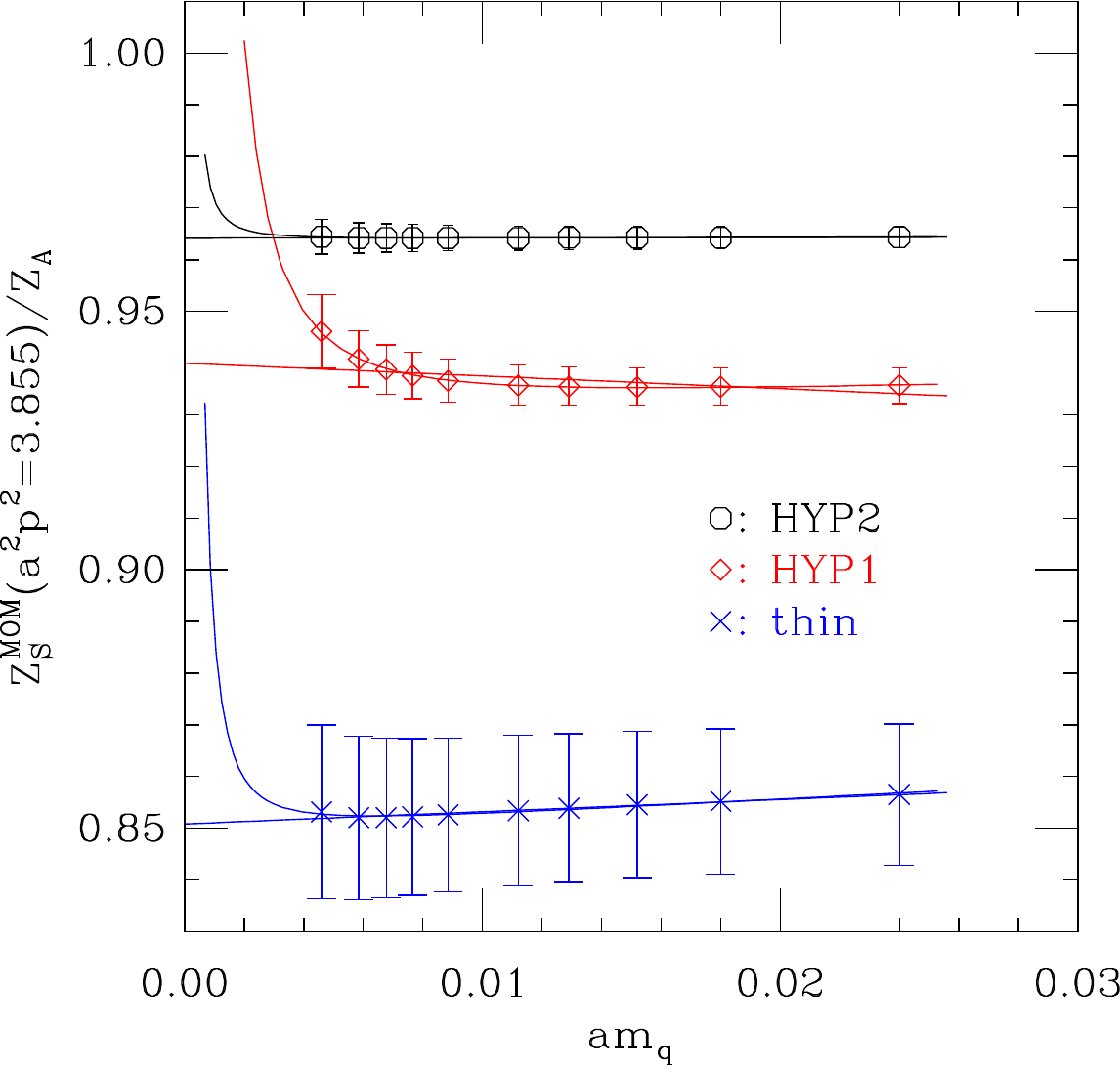}
\includegraphics[width=0.49\textwidth]{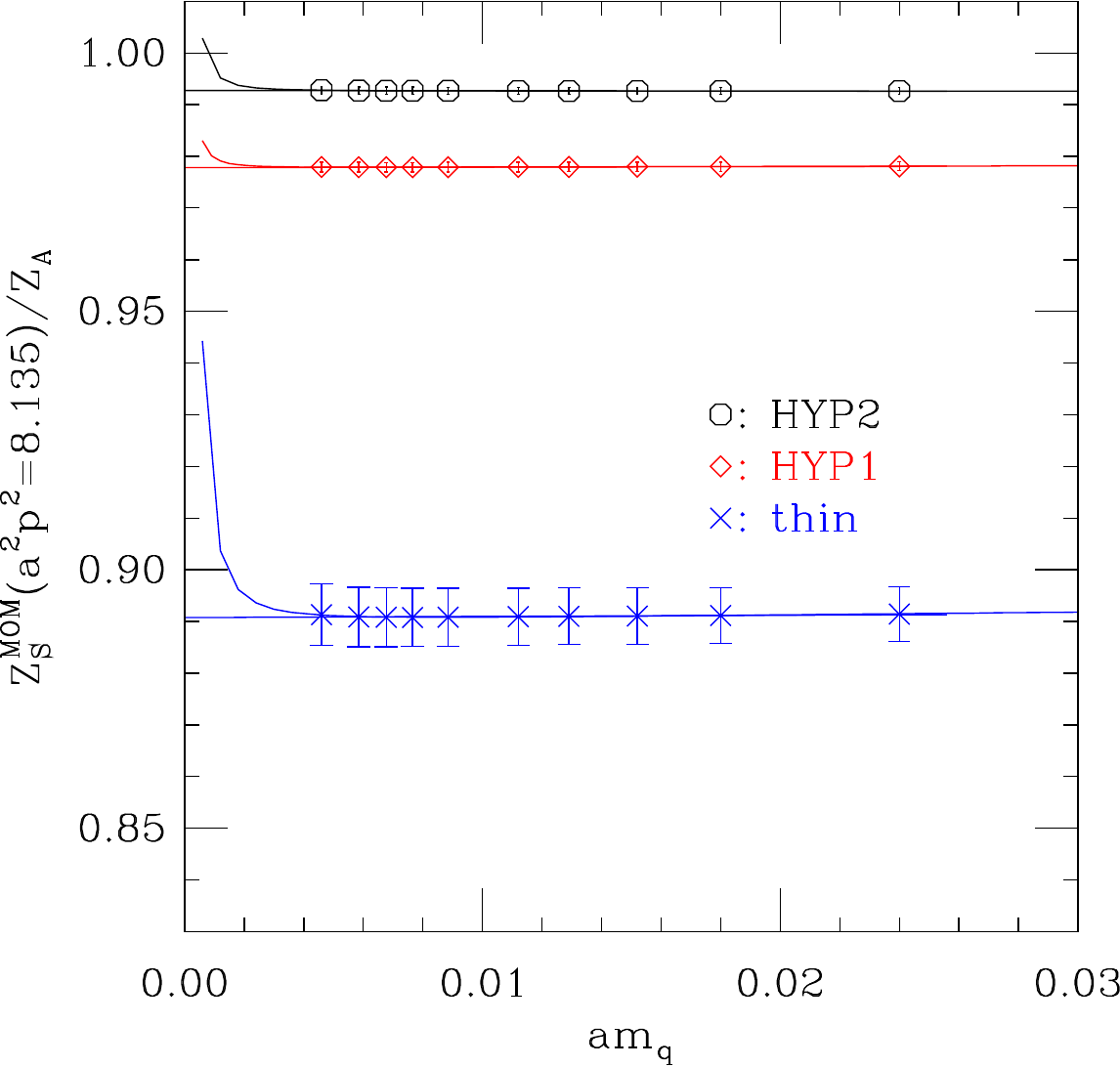}
\caption{
  Examples of chiral extrapolations of the valence quark for all three smearing cases at the scale $a^2p^2=3.855$ (left panel) or $8.135$ (right panel) on ensemble f004. Both the ansatz in Eq.(\ref{eq:zs-extrap}) and a linear function are tried for the chiral extrapolation. }
 \label{fig:zs-mom}
\end{figure}

The chiral extrapolation of the valence quark is done with an ansatz
\begin{equation}
\frac{Z_S}{Z_A}(am_q)=\frac{A_s}{(am_q)^2}+B_s+C_s\cdot(am_q),
\label{eq:zs-extrap}
\end{equation}
which was also used in Refs.~\cite{Aoki:2007xm,Bi:2017ybi,He:2022lse,Liu:2013yxz}. $B_s$ is taken as the chiral limit value of $Z_S/Z_A$. We find the contribution of the first term on
the right-hand side of Eq.~(\ref{eq:zs-extrap}) in our data is small. Thus, we also tried linear extrapolations in $am_q$ and checked that consistent results are obtained
in the chiral limit. Examples of both extrapolations at $a^2p^2=3.855$ or $8.135$ on ensemble f004 are shown in figure~\ref{fig:zs-mom}. In our following analyses we generally use data points at scales above $a^2p^2=4$ to avoid possible large nonperturbative effects.

\begin{figure}[htbp]
\centering
\includegraphics[width=0.49\textwidth]{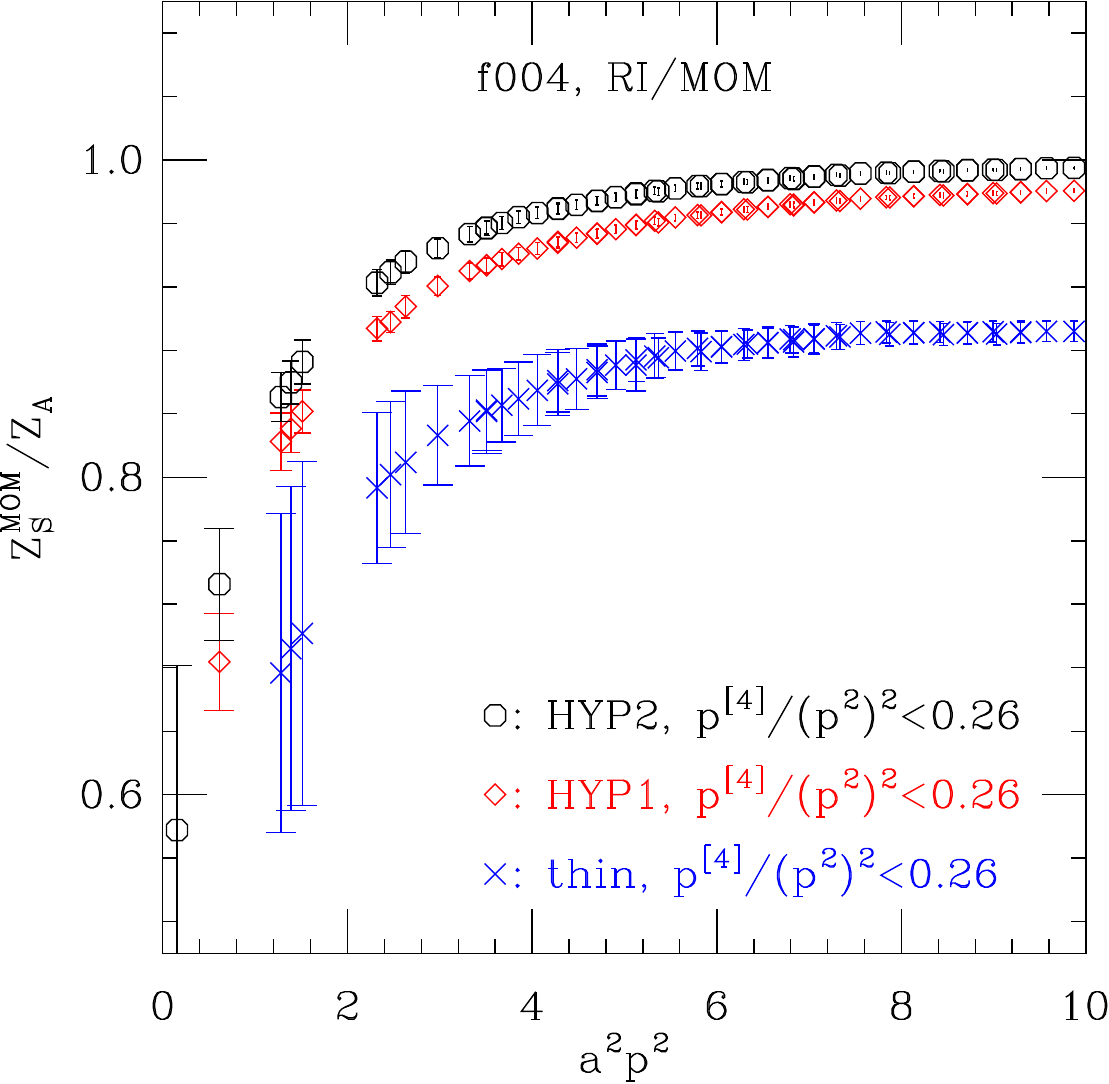}
\includegraphics[width=0.49\textwidth]{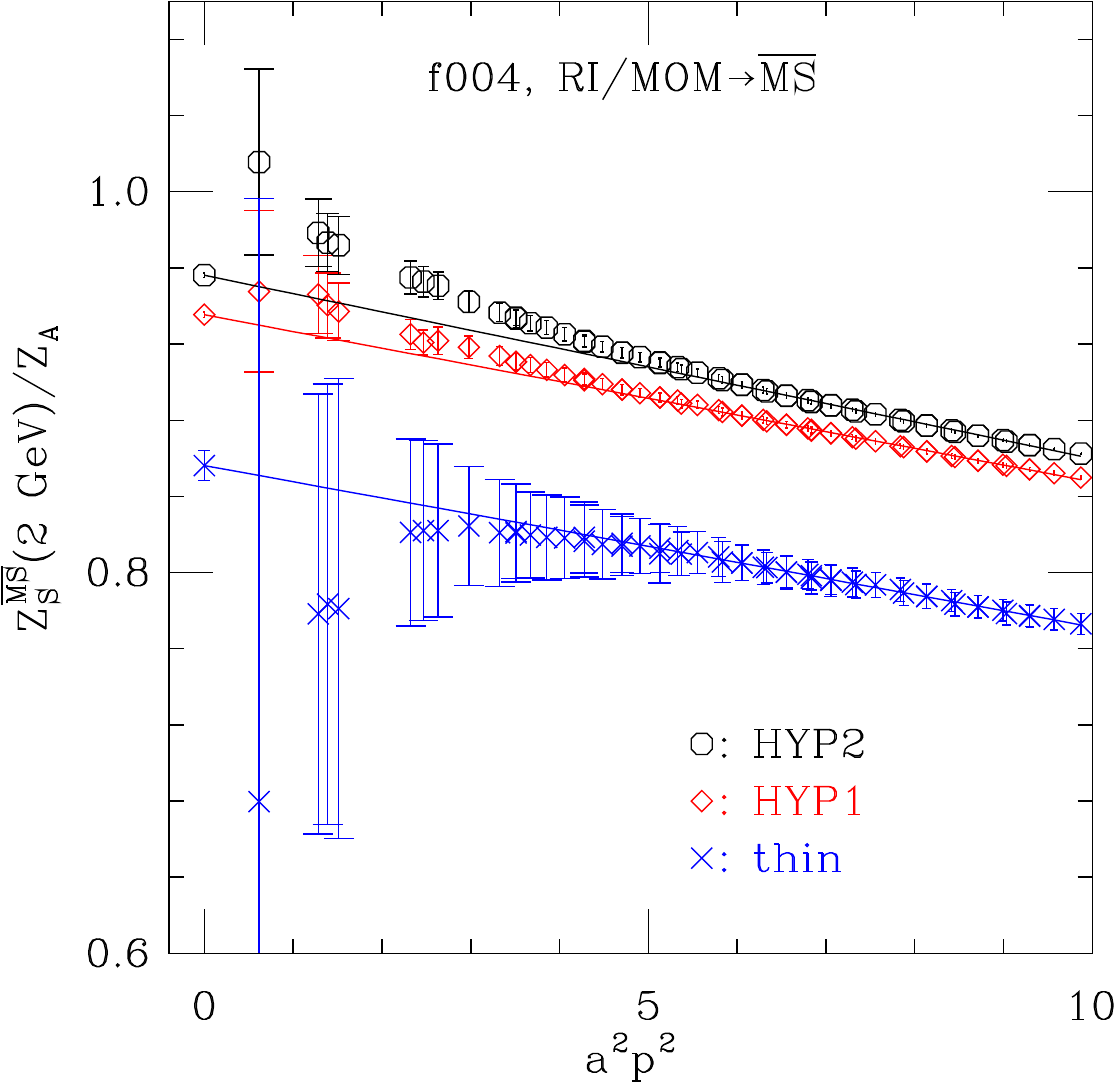}
\caption{Left panel: $Z_S^{\rm MOM}/Z_A$ as a function of the renormalization scale squared $a^2p^2$ from Eq.~(\ref{eq:zsza}) for no smearing, HYP1 and HYP2, respectively. Right panel: $Z_S^{\msbar}(2\mbox{ GeV}; a^2p^2)/Z_A$ obtained through the MOM scheme as a function of the initial scale $a^2p^2$ for the three smearing cases. The straight lines show linear extrapolations to $a^2p^2=0$ using data points in the range $a^2p^2\in[5, 10]$.}
 \label{fig:zs-mom-running}
\end{figure}

For the three smearing cases we obtain $Z_S^{\rm MOM}/Z_A^{\rm MOM}$ in the valence
quark chiral limit by using Eq.~(\ref{eq:zs-extrap}). The results are shown in the left panel of figure~\ref{fig:zs-mom-running}.

The conversion to the $\msbar$ scheme is done by using the ratio
in Landau gauge up to four loops~\cite{Franco:1998bm,Chetyrkin:1999pq,Gracey:2022vjc}
\begin{eqnarray}
\frac{Z_S^{\overline{\rm MS}}}{Z_S^{\rm MOM}}&=&
1+\frac{16}{3}\left(\frac{\alpha_s}{4\pi}\right)+\left(188.6513392
-9.888888972n_f\right)
\left(\frac{\alpha_s}{4\pi}\right)^2\nonumber \\
&&+\left(7944.242769-888.4579373n_f
+13.65785627n_f^2\right)\left(\frac{\alpha_s}{4\pi}\right)^3\nonumber\\
&&+(386340.3540-68772.60194n_f+2976.616735n_f^2-27.97607914n_f^3)\nonumber\\
&&\times\left(\frac{\alpha_s}{4\pi}\right)^4+O(\alpha_s^5),
\label{eq:ratio_zs_zp}
\end{eqnarray}
where $n_f$ is the number of flavors.
To run the results in the $\msbar$ scheme from the initial scale $|p|$ to 2 GeV, we use the quark mass anomalous dimension given
in Ref.~\cite{Chetyrkin:1999pq} since $Z_S=Z_m^{-1}$.
The four-loop (next-to-next-to-next-to-leading order) running results are shown in the right panel of
figure~\ref{fig:zs-mom-running}. Here we use the inverse lattice spacing $1/a=2.383(9)$ GeV
as determined in~\cite{RBC:2014ntl} to get the position of 2 GeV. This $1/a$ is about 4\% higher than that used in our previous work~\cite{Liu:2013yxz}. $\Lambda_{\rm QCD}^\msbar=332(17)$ MeV~\cite{ParticleDataGroup:2016lqr} is used to calculate the strong coupling
constant by using its perturbative running to four loops~\cite{Alekseev:2002zn,vanRitbergen:1997va}.

\begin{figure}[htbp]
\centering
\includegraphics[width=0.49\textwidth]{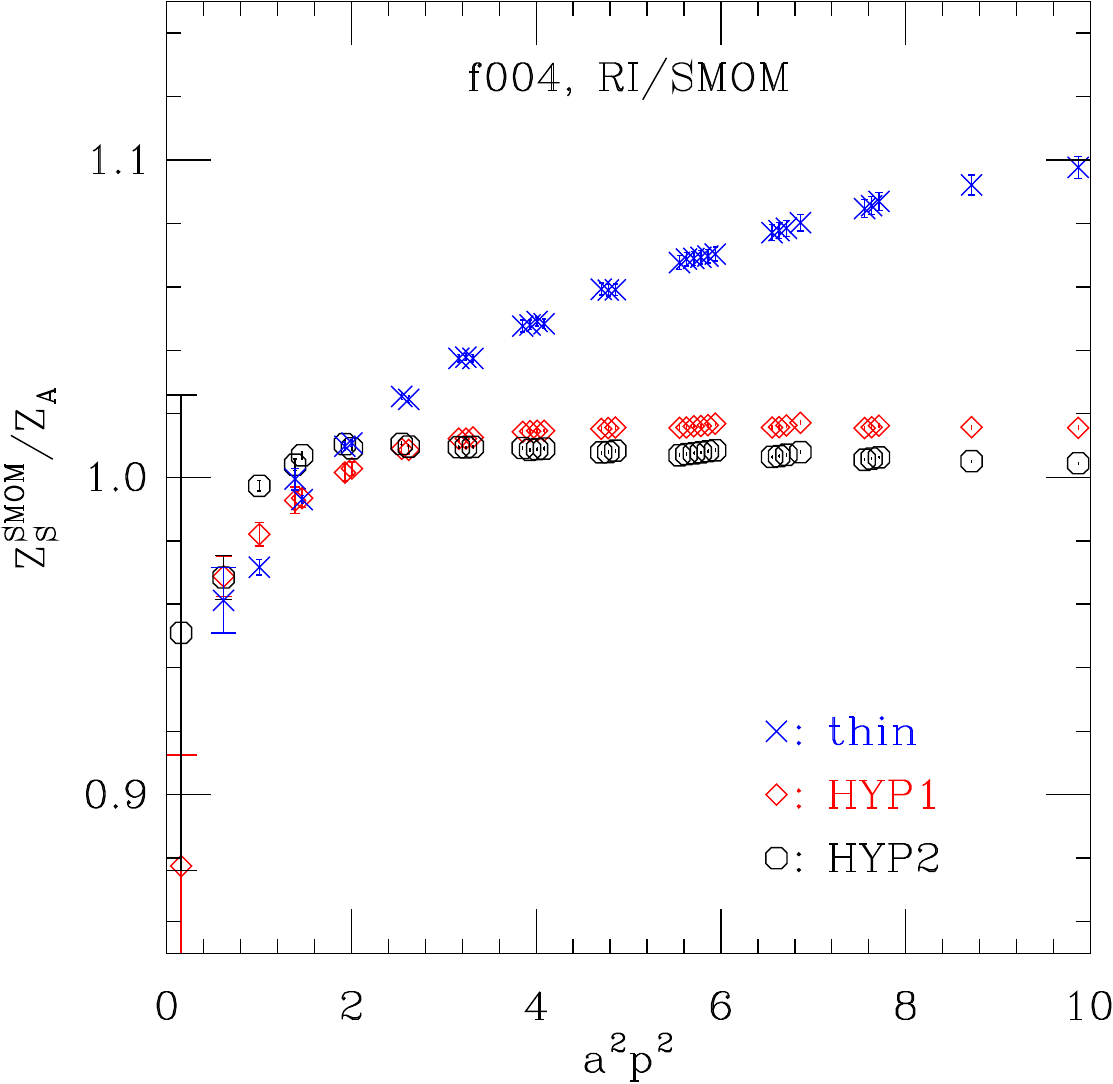}
\includegraphics[width=0.49\textwidth]{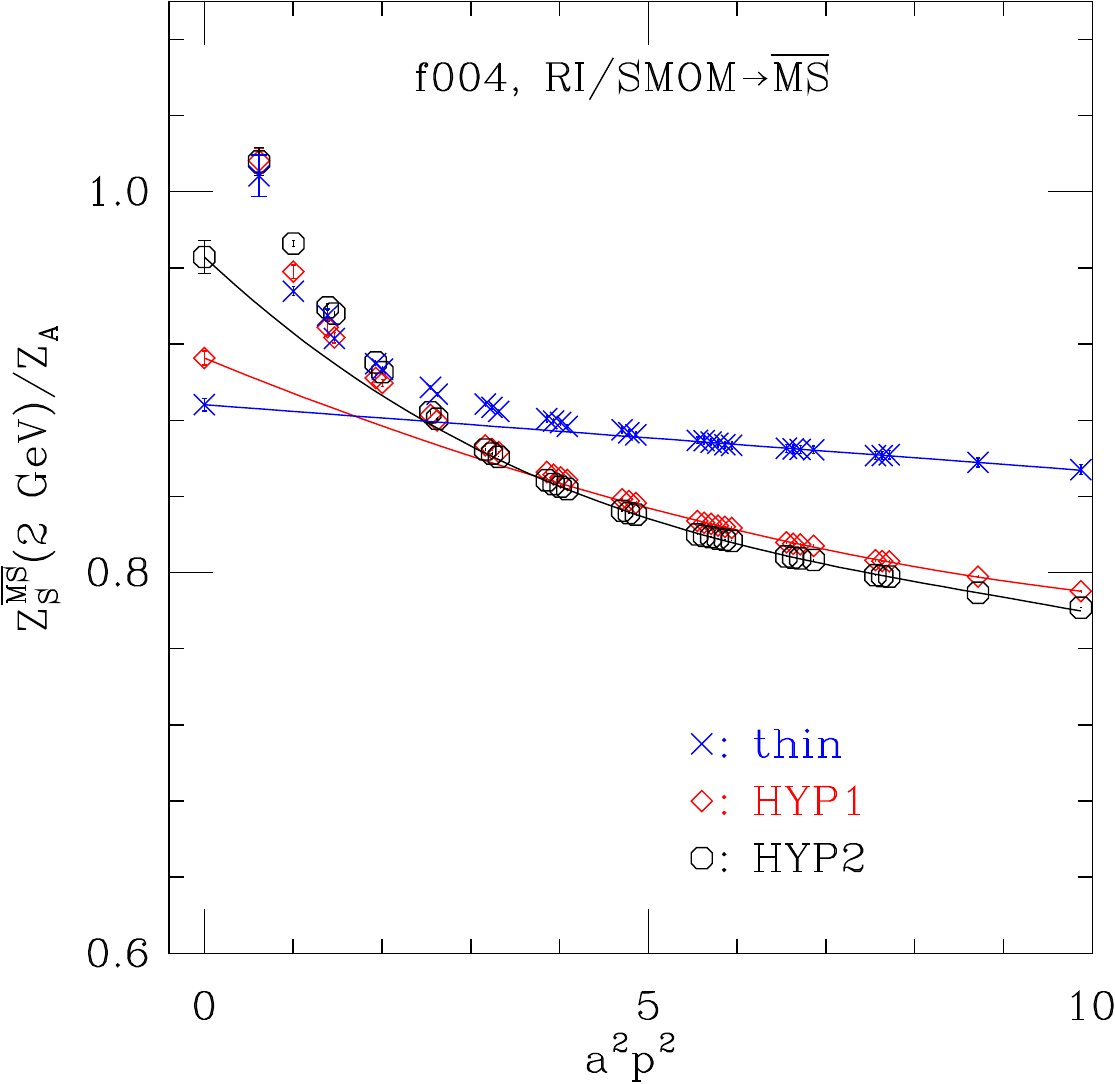}
\caption{Left panel: $Z_S^{\rm SMOM}(a^2p^2)/Z_A$ as a function of the renormalization scale $a^2p^2$ for the three smearing cases. Right panel:
$Z_S^{\msbar}(2\mbox{ GeV}; a^2p^2)/Z_A$ obtained through SMOM scheme as a function of the initial scale $a^2p^2$ for the three smearing cases.
The lines are polynomial extrapolations to $a^2p^2=0$ as explained in the text.
}
 \label{fig:zs-smom}
\end{figure}

Now we turn to the SMOM scheme. The ratio
\begin{equation}
\frac{Z_S^{\rm SMOM}}{Z_A^{\rm SMOM}}=\left.\frac{\Gamma_A(p_1,p_2)}{\Gamma_S(p_1,p_2)}\right|_{\rm sym}
\end{equation}
for the three smearing cases is shown in the left panel of figure~\ref{fig:zs-smom} in the valence quark massless limit.
Here
\begin{equation}
\Gamma_A(p_1,p_2)=\frac{1}{12q^2}\Tr[q_\mu\Lambda^\mu_{A,B}(p_1,p_2)\gamma_5 q\!\!\!/],\quad\quad
\Gamma_S(p_1,p_2)=\frac{1}{12}\Tr[\Lambda_{S,B}(p_1,p_2)].
\end{equation}
The massless limit is obtained from linear extrapolations in the valence quark mass.

The perturbative conversion ratio to the $\msbar$ scheme has been calculated
up to three loops~\cite{Gorbahn:2010bf,Almeida:2010ns,Kniehl:2020sgo,Bednyakov:2020ugu}
\begin{eqnarray}
\frac{Z_S^{\rm RI/SMOM}}{Z_S^{\msbar}}&=&1-0.6455188560\left(\frac{\alpha_s}{4\pi}\right)\nonumber\\
&&-(22.60768757-4.013539470n_f)\left(\frac{\alpha_s}{4\pi}\right)^2\nonumber\\
&&-(860.2874030-164.7423004n_f+2.184402262n_f^2)\left(\frac{\alpha_s}{4\pi}\right)^3.
\end{eqnarray}
After the conversion we can run $Z_S^{\msbar}(a^2p^2)/Z_A$ to 2 GeV. The results are shown
in the right panel of figure~\ref{fig:zs-smom}.

\subsubsection{Comparison of the three smearing cases (via MOM)}
As shown in the right panel of figure~\ref{fig:zs-mom-running}, the dependence of $Z_S^{\msbar}(2\mbox{ GeV}, a^2p^2)/Z_A$ on $a^2p^2$ is similar for all three
cases when MOM is used as the intermediate scheme.
At large momentum scale this $a^2p^2$ dependence can be described by a polynomial function
\begin{equation}
\frac{Z_S^{\msbar}(2\mbox{ GeV}, a^2p^2)}{Z_A}=\frac{Z_S^{\msbar}(2\mbox{ GeV})}{Z_A}+c_1(a^2p^2)+c_2(a^2p^2)^2,
\label{eq:a2p2-effects}
\end{equation}
where the last two terms on the right-hand side contain the discretization effects in
$Z_S^{\msbar}(2\mbox{ GeV}, a^2p^2)/Z_A$. The fittings of Eq.~(\ref{eq:a2p2-effects}) without
the $c_2$ term to data
points in the range $a^2p^2\in [5, 10]$ give $\chi^2/\mbox{dof}<1$ for all three smearing cases.
Therefore the $\mathcal{O}((a^2p^2)^2)$ discretization effects are smaller than our
current statistical error. We take this linear extrapolation value $Z_S^{\msbar}(2\mbox{ GeV})/Z_A$ as the final result, which is collected in table~\ref{tab:zs-f004}.
\begin{table}
\begin{center}
\caption{$Z_S^{\msbar}(2\mbox{ GeV})/Z_A$ on ensemble f004 for the three smearing
cases. These values are obtained after removing the discretization effects by using
a straight-line extrapolation in $a^2p^2\in [5, 10]$ (MOM, thin of SMOM), or Eq.~(\ref{eq:a2p2-effects}) (HYP1 of SMOM) or a third order polynomial (HYP2 of SMOM).
The error is for statistics and $a^2p^2$ extrapolation. }
\begin{tabular}{cccc}
\hline\hline
$Z_S^{\msbar}(2\mbox{ GeV})/Z_A$ & thin & HYP1 & HYP2 \\
\hline
MOM & 0.8561(79) & 0.9353(15) & 0.9561(12) \\
SMOM & 0.8881(32) & 0.9126(37) & 0.9657(88) \\
\hline\hline
\end{tabular}
\label{tab:zs-f004}
\end{center}
\end{table}
The fitting range of $a^2p^2$ is varied to $[4, 10]$ and $[6, 10]$ to estimate
the systematic error in $Z_S^{\msbar}(2\mbox{ GeV})/Z_A$ given
in table~\ref{tab:zs_error} from the choice of the lower bound.

There are curvatures in $Z_S^{\msbar}(2\mbox{ GeV}, a^2p^2)/Z_A$ for HYP1 and HYP2 especially at $a^2p^2<5$. Thus we also try the following ansatz to study the $a^2p^2$ dependence for these two smearing cases
\begin{equation}
\frac{Z_S^{\msbar}(2\mbox{ GeV}, a^2p^2)}{Z_A}=\frac{Z_S^{\msbar}(2\mbox{ GeV})}{Z_A}+\frac{c_{-1}}{a^2p^2}+c_1(a^2p^2),
\label{eq:zs-c-1}
\end{equation}
\begin{equation}
\frac{Z_S^{\msbar}(2\mbox{ GeV}, a^2p^2)}{Z_A}=\frac{Z_S^{\msbar}(2\mbox{ GeV})}{Z_A}+\frac{c_{-1}}{a^2p^2}+c_1(a^2p^2)+c_2(a^2p^2)^2.
\label{eq:zs-c-1-c2}
\end{equation}
We find both Eq.~(\ref{eq:a2p2-effects}) and Eq.~(\ref{eq:zs-c-1}) can describe the data well in the
range $a^2p^2\in [2, 10]$ with a roughly same $\chi^2/\mbox{dof}$, which is less than 1. Thus, it is hard to attribute the curvature to only nonperturbative effect or $\mathcal{O}(a^2p^2)$ effect. We then use Eq.~(\ref{eq:zs-c-1-c2}) to fit the data in $a^2p^2\in [5, 10]$. The resulted center values of $Z_S^{\msbar}(2\mbox{ GeV})/Z_A$ change by 0.82\% and 0.41\% for HYP1 and HYP2, respectively. This ansatz dependence is checked on all three ensembles f004, f006 and f008 for HYP1. The resulted change (0.98\%) in the final result of $Z_S^\msbar(2\mbox{ GeV})/Z_A$ in the sea quark massless limit is given in table~\ref{tab:zs_error} as a systematic error.

\subsubsection{Comparison of the three smearing cases (via SMOM)}
When the SMOM scheme is used as the intermediate scheme, the dependence of
$Z_S^{\msbar}(2\mbox{ GeV}, a^2p^2)/Z_A$ on $a^2p^2$ (right panel of figure~\ref{fig:zs-smom}) looks similar for the two smearing cases HYP1 and HYP2, which, however, seems to be quite different from the thin link case.
The $a^2p^2$ dependence is more flat at $a^2p^2>4$ when no smearing is applied.
All linear extrapolations in $a^2p^2$ in the ranges $a^2p^2\in [4, 10]$, $[5, 10]$ and $[6, 10]$
have $\chi^2/\mbox{dof}<1$ and give consistent extrapolated results for the thin link case.
The numbers from using the range $[5, 10]$ are given in table~\ref{tab:zs-f004}.

If we use Eq.~(\ref{eq:a2p2-effects}) to fit $Z_S^{\msbar}(2\mbox{ GeV}, a^2p^2)/Z_A$ with the $c_2$ term for the HYP1 case, the $\chi^2/\mbox{dof}$
are, respectively, 1.78, 0.78 and 1.06 for $a^2p^2\in [4, 10]$, $[5, 10]$ and $[6, 10]$.
If we drop the $c_2$ term, that is to say, use a linear fit, then we get large $\chi^2/\mbox{dof}$ ($42, 8.7$ and $3.7$ for the three ranges, respectively). Thus we choose the second order polynomial
extrapolation and get $Z_S^{\msbar}(2\mbox{ GeV})/Z_A=0.9126(37)$ in table~\ref{tab:zs-f004}.
The dependence on extrapolation ansatz is checked by using Eqs.~(\ref{eq:zs-c-1},\ref{eq:zs-c-1-c2}) on all three ensembles f004, f006 and f008. The variation in the center value is treated as a systematic error as collected in table~\ref{tab:zs_error}.

For the case HYP2 of SMOM we tried six ansatz: linear, second and third order polynomials of $a^2p^2$ with or without an inverse term $c_{-1}/(a^2p^2)$ to take into account possible nonperturbative effects at small $a^2p^2$.
None of them can easily fit the data in the above three ranges of $a^2p^2$.
The $\chi^2/\mbox{dof}$ are in the range 2.3 to 107.

One possible reason is the following. The statistical error decreases as the level of smearing increases. Then the other systematic errors (such as the O(4) to H(4) breaking effects) are no longer small compared with the statistical error. We use the (near) diagonal momentum modes to suppress those effects. However, in the case of SMOM more off-diagonal momentum modes have to be used than in the MOM case.
Therefore, we increase the error of the fitting results by $\sqrt{\chi^2/\rm dof}$ to compensate the ignorance of those systematic effects.

If we use the same range $a^2p^2\in [3.5, 9]$ as that in Ref.~\cite{He:2022lse} for ensemble 64I and
use a same third order polynomial,
then we obtain $\chi^2/\mbox{dof}=41.4/17=2.4$ and $Z_S^{\msbar}(2\mbox{ GeV})/Z_A=0.9657(88)$,
where the error has been enlarged by $\sqrt{\chi^2/\mbox{dof}}$.

The $a^2p^2$ dependence of $Z_S^{\msbar}(2\mbox{ GeV}, a^2p^2)/Z_A$ obtained through the SMOM scheme for the case HYP2 cannot be readily described by polynomial discretization effects even at large $a^2p^2$. This dependence seems to change significantly as one applies smearing compared with the thin link case.
For the case HYP1 the extrapolated results are more sensitive to the fitting range of $a^2p^2$
as shown by the corresponding systematic uncertainty in table~\ref{tab:zs_error} compared with MOM.
When MOM is used as the intermediate scheme, the $a^2p^2$ dependence of $Z_S^{\msbar}(2\mbox{ GeV}, a^2p^2)/Z_A$ can be well described by a linear function for
all three smearing cases at large $a^2p^2$.
Thus we use the results obtained through the MOM scheme as our final results for $Z_S^{\msbar}$.

\subsubsection{Final results for one level of HYP smearing}
On the other two ensembles f006 and f008 with smearing HYP1, we do similar calculations in the
MOM scheme and obtain $Z_S^{\msbar}(2\mbox{ GeV})/Z_A=0.9287(26)$ and $0.9352(17)$, respectively, in the valence quark massless limit. The results on all three ensembles
are collected in table~\ref{tab:zs-hyp1} as well as those obtained through the SMOM scheme.
\begin{table}
\begin{center}
\caption{$Z_S^{\msbar}(2\mbox{ GeV})/Z_A$ on ensembles f004, f006 and f008 for HYP1. They are obtained after removing the discretization effects by using
a straight-line (via MOM) or a second-order polynomial (via SMOM) extrapolation in $a^2p^2$ . The error is from statistics and $a^2p^2$ extrapolation. The values in the sea quark massless limit are from linear
extrapolations. For the intermediate MOM scheme the error is enlarged by $\sqrt{\chi^2/{\rm dof}}=\sqrt{5.3}$.}
\begin{tabular}{ccccc}
\hline\hline
$Z_S^{\msbar}(2\mbox{ GeV})/Z_A$ & f004 & f006 & f008 & $am_l+am_{\rm res}=0$\\
\hline
MOM  & 0.9353(15) & 0.9287(26) & 0.9352(17) & 0.9348(88) \\
SMOM & 0.9126(37) & 0.9123(54) & 0.9102(42) & 0.9155(94) \\
\hline\hline
\end{tabular}
\label{tab:zs-hyp1}
\end{center}
\end{table}
Then with the results on all three ensembles we do a linear extrapolation in $(am_l+am_{\rm res})$ to the light sea quark massless limit and find $Z_S^{\msbar}(2\mbox{ GeV})/Z_A=0.9348(88)$ with a confidence level 0.021, where the error has been enlarged by
$\sqrt{\chi^2/{\rm dof}}=\sqrt{5.3}$. The slope from this
extrapolation is consistent with zero ($-0.09(57)$), of which the center value is used to estimate the systematic uncertainty due to a nonzero strange sea quark mass given in table~\ref{tab:zs_error}. We also tried a constant fit of $Z_S^{\msbar}(2\mbox{ GeV})/Z_A$ on the three ensembles to go to the sea quark massless limit.
The difference in the center values from the linear and constant fits is set to be the systematic error for extrapolation in $m_l$ in table~\ref{tab:zs_error}.

We list all systematic uncertainties in table~\ref{tab:zs_error} for $Z_S^{\msbar}(2\mbox{ GeV})/Z_A$. They
are estimated in similar ways as in Refs.~\cite{Liu:2013yxz,Bi:2017ybi,He:2022lse}.
The truncation error in the conversion ratio from MOM to $\msbar$ is found to be
 0.67\%, which is smaller than the 1.5\% in~\cite{Liu:2013yxz} and the 2.29\% in~\cite{He:2022lse}. One reason is that here the lower limit of $|p|$ used in the $a^2p^2$ extrapolation is about 5.3 GeV, which is higher than the 4 GeV in~\cite{Liu:2013yxz} and the 3 GeV in~\cite{He:2022lse}.
 The other reason is now we use the newly calculated four-loop conversion ratio Eq.~(\ref{eq:ratio_zs_zp}).
 Each of $\Lambda_{\rm QCD}^\msbar$ and the inverse lattice spacing is varied in one sigma to
check the resulted change in $Z_S^{\msbar}(2\mbox{ GeV})/Z_A$. The perturbative running in
the $\msbar$ scheme includes a four-loop result and thus gives a negligible truncation error.
In total the systematic uncertainty is found to be 1.28\% when one uses the MOM scheme.
\begin{table}
\begin{center}
\caption{Systematic uncertainties of $Z_S^\msbar/Z_A(2$ GeV) in the chiral limit through the MOM or SMOM scheme for case HYP1.}
\begin{tabular}{lcc}
\hline\hline
Source & Via MOM (\%) & Via SMOM (\%)  \\
\hline
Conversion ratio  & 0.67 & 0.12 \\
$\Lambda_{\rm QCD}^\msbar$  & 0.12 & 0.42 \\
Lattice spacing & 0.11 & 0.07 \\
Perturbative running &  $<0.02$ & $<0.02$ \\
Fit range of $a^2p^2$ & 0.39 & 1.10 \\
Different fit ansatz & 0.98 & 4.95 \\
$m_s^{\rm sea}\neq0$ & 0.21 & 0.98 \\
Extrapolation in $m_l$ & 0.06 & 0.42\\
Total sys. uncertainty & 1.28 & 5.20 \\
\hline\hline
\end{tabular}
\label{tab:zs_error}
\end{center}
\end{table}

Thus, our final result is $Z_S^{\msbar}(2\mbox{ GeV})/Z_A=0.9348(88)(120)$.
Adding up quadratically the statistical and systematic uncertainties, we get
\begin{equation}
Z_S^{\msbar}(2\mbox{ GeV})/Z_A=0.935(15).
\end{equation}
By using $Z_A=1.0789(10)$ from section~\ref{sec:zawi}, we then find $Z_S^{\msbar}(2\mbox{ GeV})=1.009(16)$. This center value is about 4\% (or $1.6\sigma$) away from our previous result $1.056(6)(24)$ given in~\cite{Liu:2013yxz}.
This change is mainly from the different inverse lattice spacings used here and there.
Also, the $a^2p^2$ extrapolation ranges are different in the two works. We confirmed that
we obtain consistent results if the same inverse lattice spacing and $a^2p^2$ range are
used here as in~\cite{Liu:2013yxz}. Through the SMOM scheme one gets $Z_S^{\msbar}(2\mbox{ GeV})=0.988(53)$, which agrees with that obtained through the MOM scheme.
Our final result $Z_S^{\msbar}(2\mbox{ GeV})=1.009(16)$ agrees
with the $1.034(25)$ given in~\cite{He:2022lse} for its ensemble 64I, which has a similar setup as
the ensemble used here.

\subsection{Tensor current}
\begin{figure}[htbp]
\centering
\includegraphics[width=0.49\textwidth]{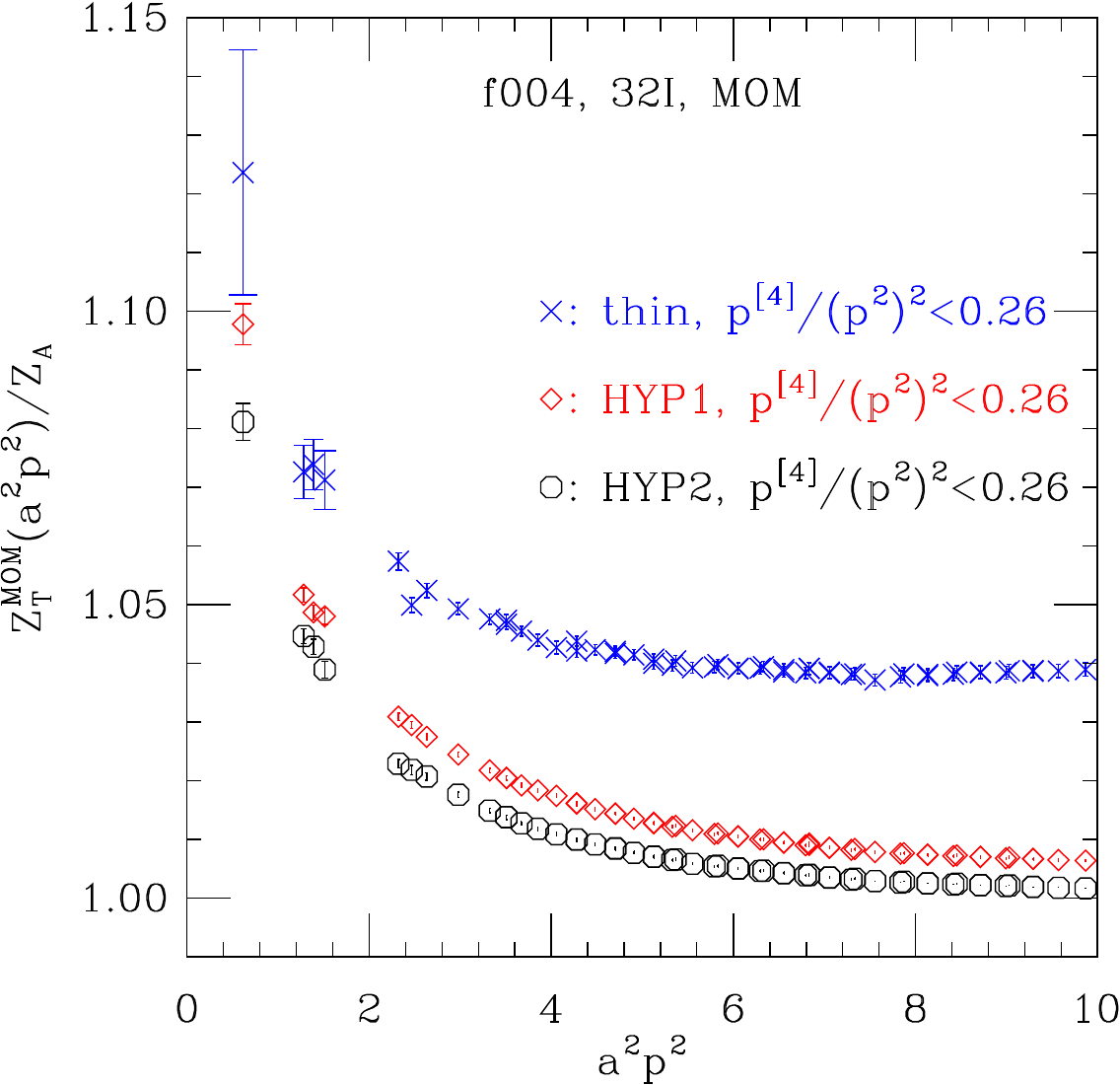}
\includegraphics[width=0.49\textwidth]{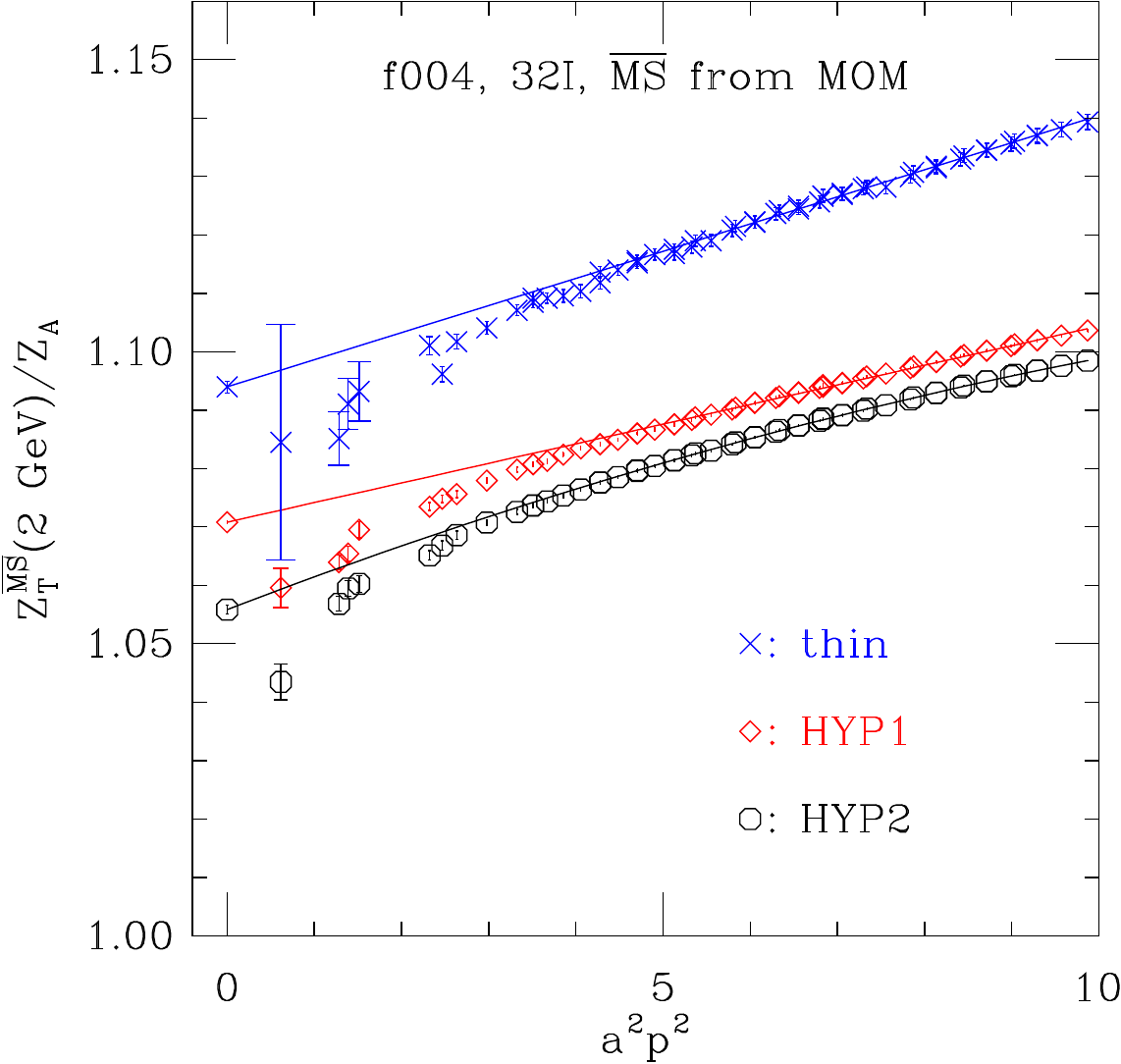}
\caption{Left panel: $Z_T^{\rm MOM}(a^2p^2)/Z_A$ as a function of the renormalization scale $a^2p^2$ for the three smearing cases on ensemble f004. Right panel:
  $Z_T^{\msbar}(2\mbox{ GeV}; a^2p^2)/Z_A$ as a function of the initial scale $a^2p^2$ for the three smearing cases obtained through the MOM scheme.}
 \label{fig:zt-mom}
\end{figure}

The renormalization constant of the tensor current is needed in calculating
observables such as the tensor decay constant of vector mesons~\cite{Chen:2020qma} and nucleon isovector tensor charge~\cite{Horkel:2020hpi}. The ratio of the RCs of the tensor
and axial vector current in the MOM scheme is given by
\begin{equation}
\frac{Z_T^{\rm MOM}}{Z_{A}^{\rm MOM}}=\left.\frac{\Gamma_A(p)}{\Gamma_T(p)}\right|_{p^2=\mu^2},
\label{eq:zt_mom}
\end{equation}
where
\begin{equation}
\Gamma_T(p)=\frac{1}{144}\Tr[\Lambda_{T,B}^{\mu\nu}(p)\sigma_{\mu\nu}].
\end{equation}
The left panel of figure~\ref{fig:zt-mom} shows the numerical results of the above ratio for the three smearing cases on ensemble f004 in the linearly extrapolated valence quark massless limit.

The conversion ratio to the $\msbar$ scheme for $Z_T^{\rm MOM}$ up to four loops is given by~\cite{Gracey:2003yr,Chetyrkin:1999pq,Gracey:2022vqr}
\begin{eqnarray}
\frac{Z_T^\msbar}{Z_T^{\rm MOM}}
&=&1+\left(-35.49868825+3.197530250n_f\right)\left(\frac{\alpha_s}{4\pi}\right)^2\nonumber\\
&&+\left(-1516.369372+235.7846995n_f-4.837941n_f^2\right)
\left(\frac{\alpha_s}{4\pi}\right)^3\nonumber\\
&&+(-62979.85943+15747.71519n_f-879.7527617n_f^2+9.395037500n_f^3)\nonumber\\
&&\times\left(\frac{\alpha_s}{4\pi}\right)^4+\mathcal{O}(\alpha_s^5).
\label{eq:ztri2msbar}
\end{eqnarray}
After finishing the conversion and four-loop running to 2 GeV from the initial
scale $a^2p^2$ in the $\msbar$ scheme, we obtain the right panel of figure~\ref{fig:zt-mom}.

\begin{figure}[htbp]
\centering
\includegraphics[width=0.49\textwidth]{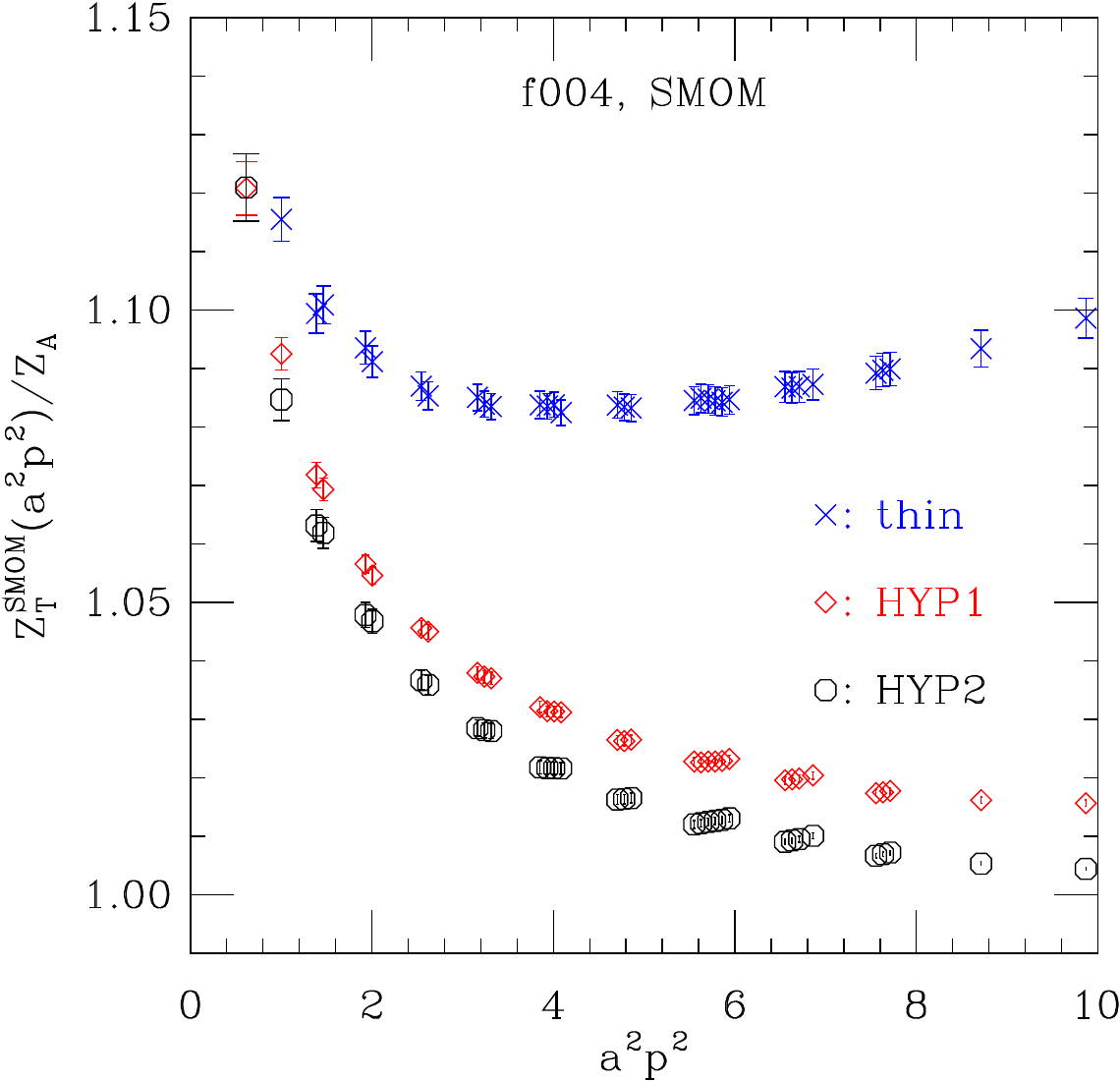}
\includegraphics[width=0.49\textwidth]{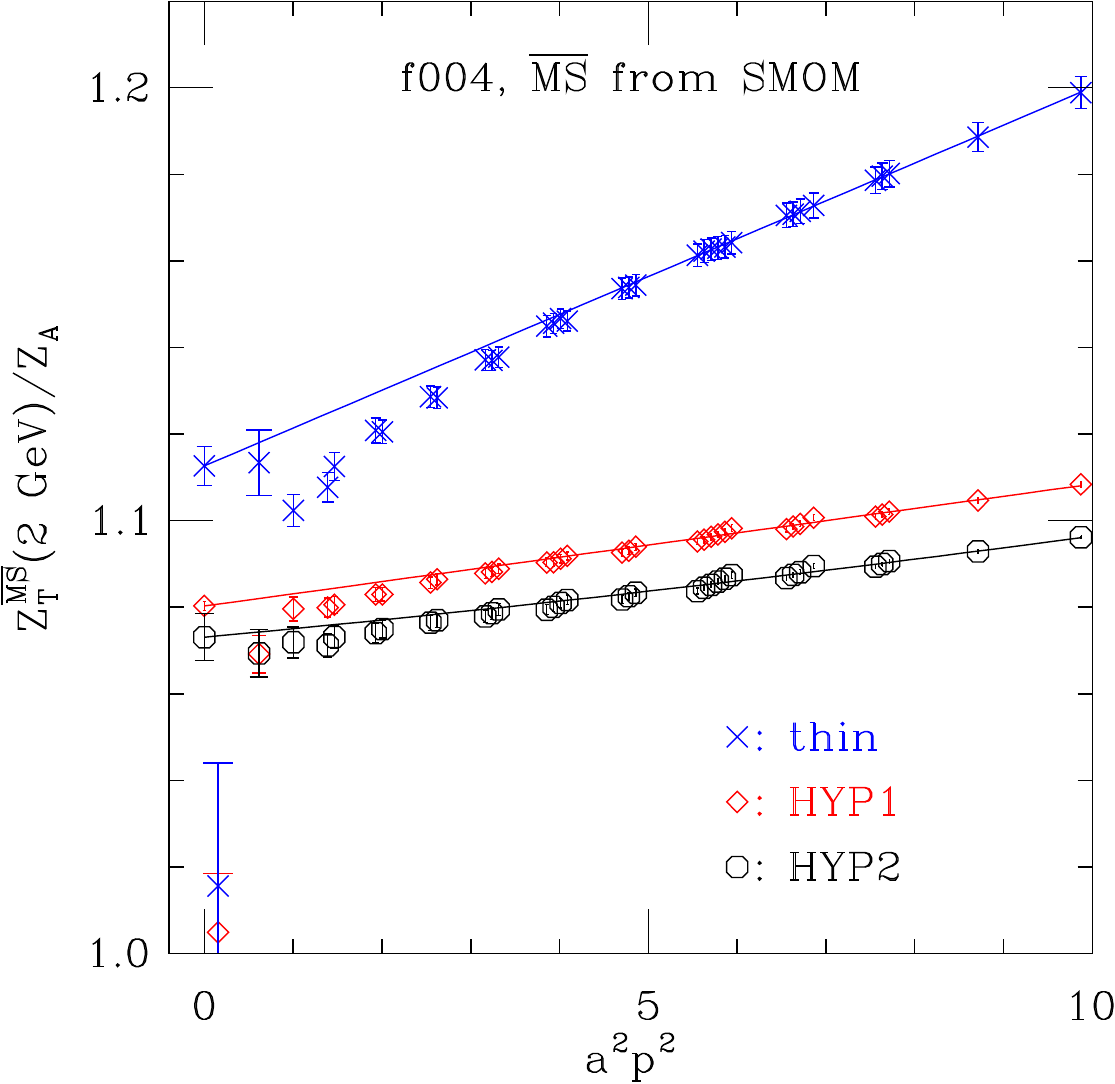}
\caption{Left panel: $Z_T^{\rm SMOM}(a^2p^2)/Z_A$ as a function of the renormalization scale $a^2p^2$ for the three smearing cases on ensemble f004. Right panel:
  $Z_T^{\msbar}(2\mbox{ GeV}; a^2p^2)/Z_A$ as a function of the initial scale $a^2p^2$ for the three smearing cases obtained through the SMOM scheme.}
 \label{fig:zt-smom}
\end{figure}

The calculation in the SMOM scheme starts with the ratio
\begin{equation}
\frac{Z_T^{\rm SMOM}}{Z_A^{\rm SMOM}}=\left.\frac{\Gamma_A(p_1,p_2)}{\Gamma_T(p_1,p_2)}\right|_{\rm sym},
\end{equation}
where
\begin{equation}
\Gamma_T(p_1,p_2)=\frac{1}{144}\Tr[\Lambda_{T,B}^{\mu\nu}(p_1,p_2)\sigma_{\mu\nu}].
\end{equation}
The numerical results of this ratio are shown in the left panel of figure~\ref{fig:zt-smom} for all three smearing cases in the valence quark massless limit on ensemble f004.
The valence chiral extrapolation is done by using a linear function in $am_q$.

The three-loop conversion of $Z_T$ from the SMOM scheme to the $\msbar$ scheme is~\cite{Kniehl:2020sgo}
\begin{eqnarray}
\frac{Z_T^{\msbar}}{Z_T^{\rm RI/SMOM}}&=&1-0.21517295\left(\frac{\alpha_s}{4\pi}\right)\nonumber\\
&&-(43.38395007-4.10327859n_f)\left(\frac{\alpha_s}{4\pi}\right)^2\nonumber\\
&&-(1950.76(11)-309.8285(28)n_f+7.063585(58)n_f^2)\left(\frac{\alpha_s}{4\pi}\right)^3\nonumber\\
&&+\mathcal{O}(\alpha_s^4).
\label{eq:zt_smom}
\end{eqnarray}
By using this conversion ratio and the anomalous dimension of $Z_T$ up to four loops in the $\msbar$ scheme~\cite{Baikov:2006ai}
we obtain $Z_T^\msbar(2\mbox{ GeV}; a^2p^2)/Z_A$ as a function of the initial scale $a^2p^2$.
The results are shown in the right panel of figure~\ref{fig:zt-smom}.

For $n_f=3$ and scale $|p|=5.3$ GeV, the conversion ratio in Eq.~(\ref{eq:zt_smom}) is
\begin{eqnarray}
&&\frac{Z_T^{\msbar}}{Z_T^{\rm RI/SMOM}}(|p|=5.3\mbox{ GeV},n_f=3)\nonumber\\
&=&1-0.017123\alpha_s-0.196779\alpha_s^2-0.546687\alpha_s^3+\mathcal{O}(\alpha_s^4)\nonumber\\
&=&1-0.0034-0.0078-0.0043+\mathcal{O}(\alpha_s^4).
\end{eqnarray}
Assuming the coefficient in the $\mathcal{O}(\alpha_s^4)$ term is 3 times as big as
that in the $\mathcal{O}(\alpha_s^3)$ term, we can estimate the size of
the $\mathcal{O}(\alpha_s^4)$ term to be around 0.0026.
Therefore, the truncation error is about 0.26\%, which is a little bit larger than
that for the MOM scheme at the same order.

One can then extrapolate $Z_T^\msbar(2\mbox{ GeV}; a^2p^2)/Z_A$ at large $a^2p^2$ to $a^2p^2=0$ to remove
lattice artifacts proportional to $a^2p^2$ (and $(a^2p^2)^2$) by using a linear function
(or a second-order polynomial) of $a^2p^2$.
To avoid the nonperturbative effects at small $a^2p^2$ we use $Z_T^\msbar(2\mbox{ GeV}; a^2p^2)/Z_A$ in the range $a^2p^2\in [5, 10]$. We vary the range to $[4, 10]$ and $[6, 10]$
to estimate the systematic error from the choice of fitting ranges.

\begin{figure}[htbp]
\centering
\includegraphics[width=0.49\textwidth]{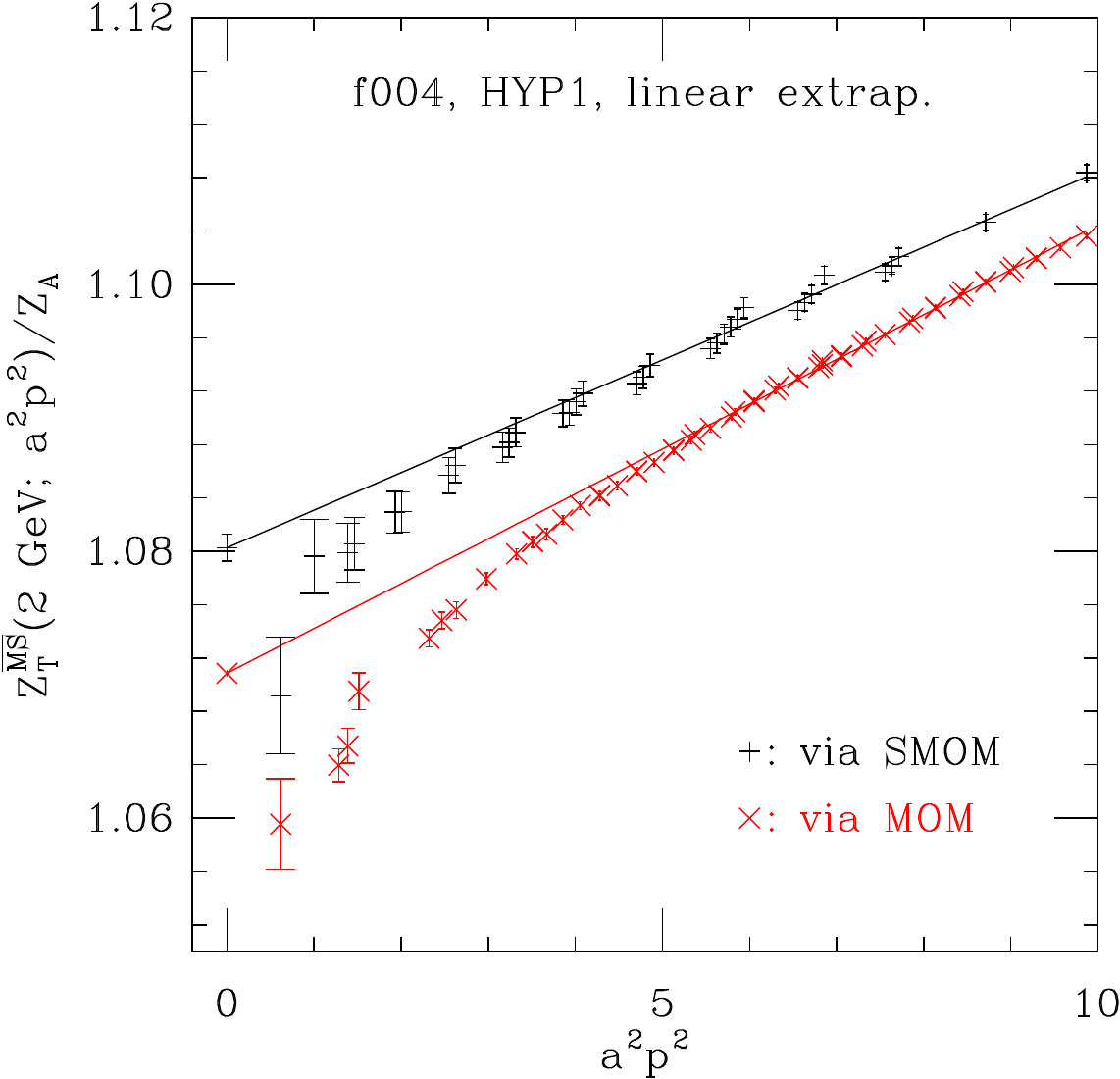}
\caption{Linear extrapolations of $Z_T^{\msbar}(2\mbox{ GeV}; a^2p^2)/Z_A$ to $a^2p^2=0$ from the two intermediate schemes. $a^2p^2$ is the initial renormalization scale squared.}
 \label{fig:zt-extrap}
\end{figure}

For the case HYP1 we find that a linear function of $a^2p^2$ is good enough ($\chi^2/\mbox{dof}<1$) to describe $Z_T^\msbar(2\mbox{ GeV}; a^2p^2)/Z_A$ in the range
$a^2p^2\in[5, 10]$ or $[6, 10]$ obtained through
either the MOM or the SMOM scheme. Figure~\ref{fig:zt-extrap} shows the linear extrapolations using the data at
$a^2p^2\in [5, 10]$ for both intermediate schemes.
For the linear extrapolations in the range
$a^2p^2\in[4, 10]$ we get $\chi^2/\mbox{dof}=1.7$, which is more or less acceptable, and $0.7$ for MOM and SMOM, respectively. The extrapolated results are given in table~\ref{tab:zt-extrap} along with the $\chi^2/\mbox{dof}$ for different fitting ranges.
The uncertainties in the table are from statistics and the extrapolations.
\begin{table}
\begin{center}
\caption{Linear extrapolations in $a^2p^2$ for $Z_T^\msbar(2\mbox{ GeV}; a^2p^2)/Z_A$ for
the two cases thin and HYP1 on ensemble f004.}
\begin{tabular}{ccccc}
\hline\hline
Thin & Fitting range & $[4, 10]$ & $[5, 10]$ & [6, 10] \\
MOM & $\chi^2/\mbox{dof}$ & 12.9/40 & 6.2/33 & 2.6/26 \\
& $Z_T^\msbar(2\mbox{ GeV})/Z_A$ & 1.0928(7) & 1.0940(10) & 1.0955(15)  \\
SMOM & $\chi^2/\mbox{dof}$ & 0.8/18 & 0.2/13 & 0.03/7 \\
& $Z_T^\msbar(2\mbox{ GeV})/Z_A$ & 1.1106(27) & 1.1126(45) & 1.1139(80)  \\
\hline
HYP1 & Fitting range & $[4, 10]$ & $[5, 10]$ & [6, 10] \\
MOM & $\chi^2/\mbox{dof}$ & 69/40 & 31.3/33 & 10.0/26 \\
& $Z_T^\msbar(2\mbox{ GeV})/Z_A$ & 1.0701(2) & 1.0709(2) & 1.0717(3)  \\
SMOM & $\chi^2/\mbox{dof}$ & 11.8/18 & 10.2/13 & 5.0/7 \\
& $Z_T^\msbar(2\mbox{ GeV})/Z_A$ & 1.0798(7) & 1.0803(10) & 1.0796(16)  \\
\hline\hline
\end{tabular}
\label{tab:zt-extrap}
\end{center}
\end{table}
The change in $Z_T^\msbar(2\mbox{ GeV})/Z_A$ is around or less than
0.2\% as we vary the fitting ranges for both intermediate schemes. The difference in
$Z_T^\msbar(2\mbox{ GeV})/Z_A$ from the two schemes is around 0.9\%.

For the case without smearing, the linear extrapolations of
$Z_T^\msbar(2\mbox{ GeV}; a^2p^2)/Z_A$ to $a^2p^2=0$ are similar to those for HYP1.
The $\chi^2/\mbox{dof}$ are smaller because the statistical uncertainties in the data are bigger. The difference in
$Z_T^\msbar(2\mbox{ GeV})/Z_A$ from the two schemes is around 1.7\%.

For the case HYP2 we find that linear extrapolations in $a^2p^2$ give large $\chi^2/\mbox{dof}$ ($>2$) when MOM is used as the intermediate scheme. Thus a second-order polynomial in $a^2p^2$ is used for the extrapolation ($\chi^2/\mbox{dof}=16/32$ for $a^2p^2\in[5, 10]$) and we obtain $Z_T^\msbar(2\mbox{ GeV})/Z_A=1.0559(7)$.
For the intermediate SMOM scheme both linear and second-order polynomial extrapolations have
$\chi^2/\mbox{dof}\sim1.6$ and give consistent extrapolated results 1.0705(9) and 1.0731(54).
The difference in
$Z_T^\msbar(2\mbox{ GeV})/Z_A$ from the two schemes is 1.6\% from
the second-order polynomial extrapolation.

We do similar analyses of $Z_T^{\msbar}(2\mbox{ GeV})/Z_A$ for the other two ensembles f006 and f008 for the case HYP1. The final results for all three ensembles are collected in table~\ref{tab:zt}.
\begin{table}
\begin{center}
\caption{$Z_T^{\msbar}(2\mbox{ GeV})/Z_A$ on ensembles f004, f006 and f008. They are obtained after removing the discretization effects by using
a straight-line extrapolation in $a^2p^2$. The error is from statistics and $a^2p^2$ extrapolation. The values in the sea quark massless limit are from linear
extrapolations, whose errors are enlarged by $\sqrt{\chi^2/{\rm dof}}$ (if $>1$).}
\begin{tabular}{ccccc}
\hline\hline
$Z_T^{\msbar}(2\mbox{ GeV})/Z_A$ & f004 & f006 & f008 & $am_l+am_{\rm res}=0$\\
\hline
MOM  & 1.0709(02) & 1.0733(02) & 1.0706(2) & 1.0721(51) \\
SMOM & 1.0803(10) & 1.0852(16) & 1.0797(9) & 1.0821(71) \\
\hline\hline
\end{tabular}
\label{tab:zt}
\end{center}
\end{table}
The values in the sea quark massless limit are from linear extrapolations in
$am_l+am_{\rm res}$. The two numbers from the two intermediate schemes are in agreement
at $1.2\sigma$, with a 1.0\% change in the center values.

The systematic uncertainties of $Z_T^{\msbar}(2\mbox{ GeV})/Z_A$ are given in table~\ref{tab:zt_error}.
The analysis procedure is similar to that for $Z_S^{\msbar}(2\mbox{ GeV})/Z_A$.
\begin{table}
\begin{center}
\caption{Systematic uncertainties of $Z_T^\msbar/Z_A(2$ GeV) in the chiral limit through the MOM and SMOM schemes.}
\begin{tabular}{lcc}
\hline\hline
Source &  MOM (\%) &  SMOM (\%)  \\
\hline
Conversion ratio           & 0.07 & 0.26 \\
$\Lambda_{\rm QCD}^\msbar$ & 0.16 & 0.15 \\
Lattice spacing            & 0.03 & 0.04  \\
Perturbative running &  $<0.01$ & $<0.01$ \\
Fit range of $a^2p^2$      & 0.07 & 0.18 \\
Different fit ansatz       & 0.53 & 0.47 \\
$m_s^{\rm sea}\neq0$ & 0.21 & 0.27 \\
Extrapolation in $m_l$ & 0.05 & 0.12 \\
Total systematic uncertainty & 0.60 & 0.66 \\
\hline\hline
\end{tabular}
\label{tab:zt_error}
\end{center}
\end{table}
To check the model dependence of the $a^2p^2$ extrapolation we also use the following ansatz
\begin{equation}
\frac{Z_T^{\msbar}(2\mbox{ GeV}, a^2p^2)}{Z_A}=\frac{Z_T^{\msbar}(2\mbox{ GeV})}{Z_A}+\frac{c_{-1}^T}{a^2p^2}+c_1^T(a^2p^2),
\label{eq:zt-c-1}
\end{equation}
besides the linear extrapolation. The resulted difference in the chiral limit is included as
a systematic error.
Finally, one obtains
\begin{eqnarray}
Z_T^\msbar(2\mbox{ GeV})/Z_A&=&1.0721(51)(65)\mbox{ (via MOM),}\label{eq:zt-mom}\\
Z_T^\msbar(2\mbox{ GeV})/Z_A&=&1.0821(71)(71)\mbox{ (via SMOM),}\label{eq:zt-smom}
\end{eqnarray}
where the two errors are statistical and systematic, respectively.
We take the number from the MOM scheme as the final result for $Z_T^\msbar(2\mbox{ GeV})/Z_A$
and treat half of the difference in the center values of Eqs.(\ref{eq:zt-mom},\ref{eq:zt-smom}) as another systematic error
to give
\begin{equation}
Z_T^\msbar(2\mbox{ GeV})/Z_A=1.0721(51)(65)(50).
\end{equation}
Adding up all the errors quadratically and using $Z_A=1.0789(10)$ from section~\ref{sec:zawi}, we get
\begin{equation}
Z_T^\msbar(2\mbox{ GeV})=1.157(11),
\end{equation}
which is in agreement with the result 1.150(5) for ensemble 64I in Ref.~\cite{He:2022lse} (table XVII).

\subsection{Quark field renormalization}
QCD propagators such as the quark propagator can provide nonperturbative information about QCD as, for example, discussed in Ref.~\cite{Sternbeck:2016ltn}. Studies of quark propagators in Landau gauge by using lattice QCD can be found in, for example, Refs.~\cite{Blossier:2010vt,Virgili:2022wfx}.
The quark field renormalization constant $Z_q^\msbar$ in the $\msbar$ scheme is useful when quark propagators are used to
determine the quark chiral condensate as was tried in Ref.~\cite{Wang:2016lsv}.

To obtain $Z_q^\msbar$ the following ratios are calculated and then converted to the $\msbar$ scheme:
\begin{equation}
\frac{Z_q^{\rm MOM}}{Z_A^{\rm MOM}}=\Gamma_A(p)|_{p^2=\mu^2},\quad\quad\quad
\frac{Z_q^{\rm SMOM}}{Z_A^{\rm SMOM}}=\Gamma_A(p_1,p_2)|_{\rm sym}.
\end{equation}
The valence quark chiral extrapolations of those ratios are finished by using a linear
function in $am_q$ as were done in Refs.~\cite{Bi:2017ybi,He:2022lse}.

The four-loop conversion ratios to the $\msbar$ scheme for the two intermediate schemes
are~\cite{Chetyrkin:1999pq,Gracey:2022vjc}, respectively,
\begin{eqnarray}
\frac{Z_q^\msbar}{Z_q^{\text{MOM}}}&=&1+\left(-14.29753930+\frac{5}{3}n_f\right)\left(\frac{\alpha_s}{4\pi}\right)^2\nonumber\\
&&+(-945.7120794+173.2442907n_f-4.534979350n_f^2)\left(\frac{\alpha_s}{4\pi}\right)^3\nonumber\\
&&+(-44917.79393+11291.87285n_f-701.6008859n_f^2+8.854793272n_f^3)\nonumber\\
&&\times\left(\frac{\alpha_s}{4\pi}\right)^4+\mathcal{O}(\alpha_s^5),
\label{eq:zq_mom_convert}
\end{eqnarray}
and
\begin{eqnarray}
\frac{Z_q^\msbar}{Z_q^{\text{SMOM}}}&=&1+\left[-\frac{359}{9}+12\zeta_3+\frac{7}{3}n_f\right]\left(\frac{\alpha_s}{4\pi}\right)^2
  +\left[-\frac{439543}{162}+\frac{8009}{6}\zeta_3 +\frac{79}{4}\zeta_4\right.\nonumber\\
  &&\left.-\frac{1165}{3}\zeta_5+\frac{24722}{81}n_f-\frac{440}{9}\zeta_3 n_f-\frac{1570}{243}n_f^2\right]
  \left(\frac{\alpha_s}{4\pi}\right)^3\nonumber\\
  &&\left[\frac{21391}{1458}n_f^3-\frac{356864009}{5184}\zeta_5-\frac{146722043}{864}
  -\frac{29889697}{5184}\zeta_3^2-\frac{1294381}{108}\zeta_3 n_f\right.\nonumber\\
  &&\left.-\frac{1276817}{972}n_f^2-\frac{440}{9}\zeta_5n_f^2-\frac{20}{3}\zeta_4n_f^2
  +\frac{8}{27}\zeta_3n_f^3+\frac{100}{3}\zeta_6n_f+\frac{2291}{72}\zeta_4n_f\right.\nonumber\\
  &&\left.+\frac{5704}{27}\zeta_3n_f^2+\frac{565939}{864}\zeta_4+\frac{1673051}{324}\zeta_5n_f
  +\frac{3807625}{10368}\zeta_6+\frac{6747755}{288}\zeta_7\right.\nonumber\\
  &&\left.+\frac{55476671}{1944}n_f+\frac{317781451}{2592}\zeta_3
  -1029\zeta_7n_f-24\zeta_3^2n_f\right]\left(\frac{\alpha_s}{4\pi}\right)^4\nonumber\\
  &&+\mathcal{O}(\alpha_s^5).
\label{eq:zq_smom_convert}
\end{eqnarray}
From Eqs.~(\ref{eq:zq_mom_convert},\ref{eq:zq_smom_convert}) one can estimate the size of the truncated higher order terms. At $a^2p^2=5$ or $|p|=5.3$ GeV the higher order terms
are of size 0.0006 and 0.0012 for the MOM and SMOM scheme, respectively. They are given
in table~\ref{tab:zq_error} as systematic errors.

\begin{figure}[htbp]
\centering
\includegraphics[width=0.49\textwidth]{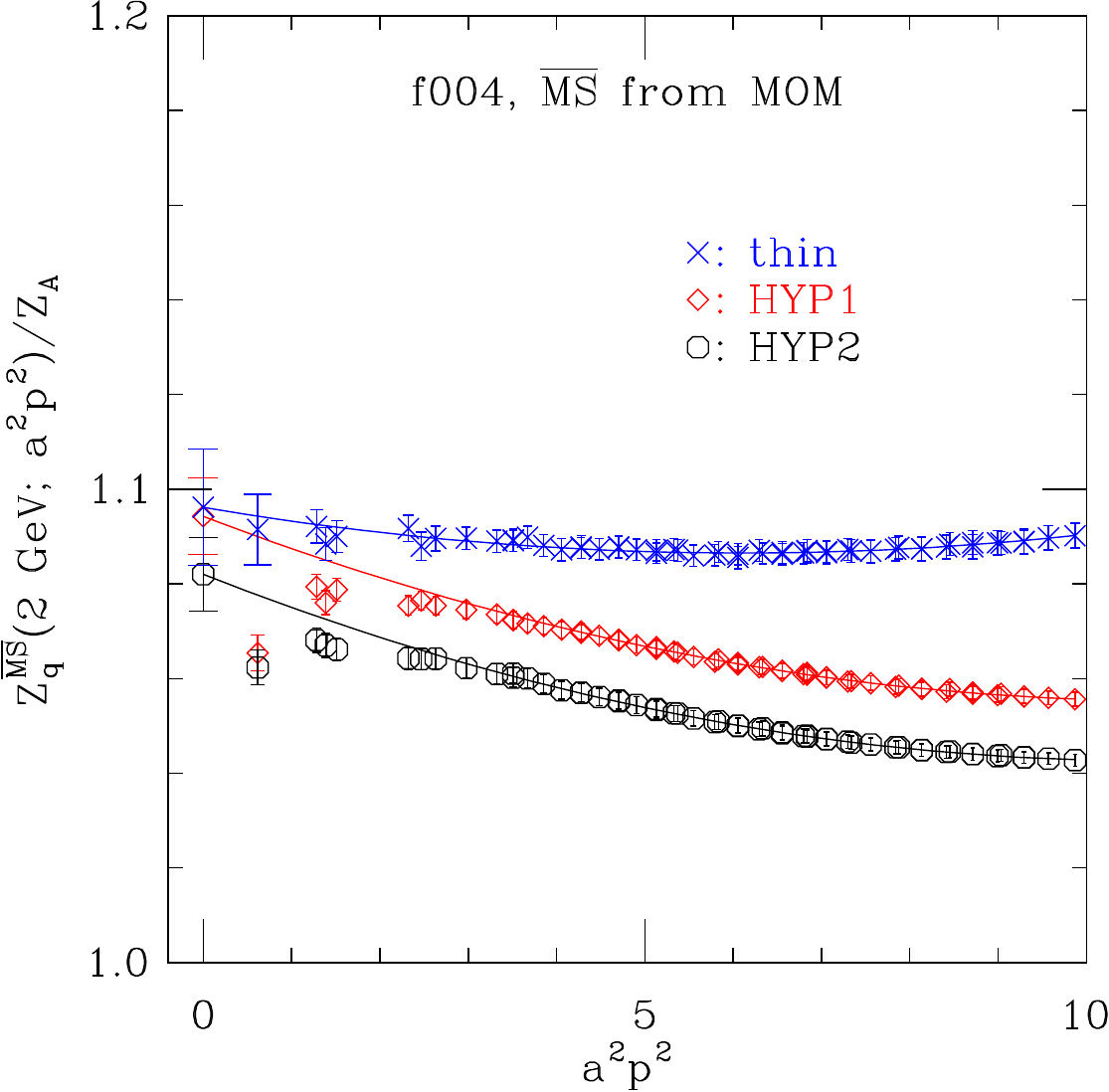}
\includegraphics[width=0.49\textwidth]{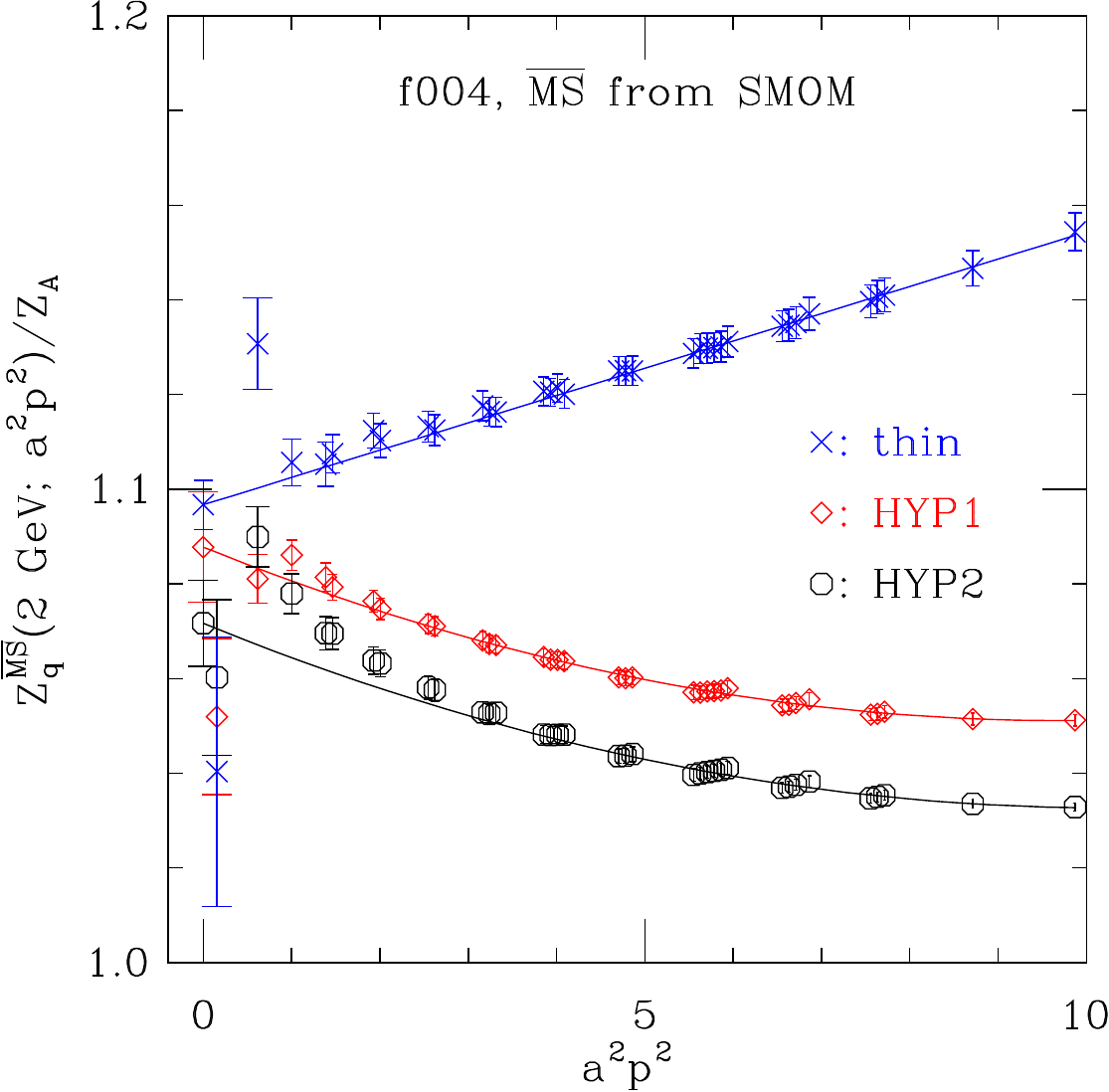}
\caption{Extrapolations of $Z_q^{\msbar}(2\mbox{ GeV}; a^2p^2)/Z_A$ to $a^2p^2=0$ from the two intermediate schemes. $a^2p^2$ is the initial renormalization scale squared.}
 \label{fig:zq}
\end{figure}

Then $Z_q^\msbar/Z_A$ at renormalization scale $a^2p^2$ can be
run to 2 GeV by using the field anomalous dimension $\gamma_q^\msbar$, which has been calculated to four loops in Landau gauge~\cite{Chetyrkin:1997dh} in perturbation theory.
The two graphs of figure~\ref{fig:zq} show $Z_q^\msbar/Z_A(2\mbox{ GeV}; a^2p^2)$ after the running as functions of the initial scale $a^2p^2$ for the two intermediate schemes, respectively.
The $a^2p^2$ dependence of
$Z_q^\msbar/Z_A(2\mbox{ GeV}; a^2p^2)$ can be described by either a leading discretization
effect proportional to $a^2p^2$ (thin of SMOM) or the following function with a higher order
term
\begin{equation}
\frac{Z_q^\msbar(2\mbox{ GeV}; a^2p^2)}{Z_A}=\frac{Z_q^\msbar(2\mbox{ GeV})}{Z_A}+
c_1^q(a^2p^2)+c_2^q(a^2p^2)^2.
\label{eq:zq-extrap}
\end{equation}
From the linear or second-order polynomial dependence on $a^2p^2$ at large scale ($a^2p^2\in[5,10]$) we extrapolate $Z_q^\msbar/Z_A(2\mbox{ GeV}; a^2p^2)$ to $a^2p^2=0$
to remove the lattice artefacts.

Besides the linear function or Eq.~(\ref{eq:zq-extrap}), we try functions with a nonperturbative term
\begin{equation}
\frac{Z_q^{\msbar}(2\mbox{ GeV}, a^2p^2)}{Z_A}=\frac{Z_q^{\msbar}(2\mbox{ GeV})}{Z_A}+\frac{c_{-1}^q}{a^2p^2}+c_1^q(a^2p^2)+c_2^q(a^2p^2)^2
\label{eq:zq-c-1-c2}
\end{equation}
to estimate the uncertainty of $Z_q^{\msbar}(2\mbox{ GeV})/Z_A$ from the fitting ansatz
as given in table~\ref{tab:zq_error}.

The extrapolated results $Z_q^\msbar/Z_A(2\mbox{ GeV})$ on the three ensembles with HYP1 are listed
in table~\ref{tab:zq} along with the values in the sea quark massless limit. The sea quark chiral
extrapolation is done linearly in $(am_l+am_{\rm res})$. A constant extrapolation in the
light sea quark mass is used
to estimate the associated systematic error. The slope from the linear extrapolation is used
to estimate the systematic error due to the nonzero strange sea quark mass.
\begin{table}
\begin{center}
\caption{$Z_q^{\msbar}(2\mbox{ GeV})/Z_A$ on ensembles f004, f006 and f008 for HYP1. They are obtained after removing the discretization effects by using
a second-order polynomial function in $a^2p^2$. The error is from statistics and $a^2p^2$ extrapolation. The values in the sea quark massless limit are from linear
extrapolations, whose error is enlarged by $\sqrt{\chi^2/{\rm dof}}=\sqrt{1.39}$ for the intermediate scheme MOM.}
\begin{tabular}{ccccc}
\hline\hline
$Z_q^{\msbar}(2\mbox{ GeV})/Z_A$ & f004 & f006 & f008 & $am_l+am_{\rm res}=0$\\
\hline
MOM &  1.094(08) & 1.110(12) & 1.095(7) & 1.097(22) \\
SMOM & 1.088(12) & 1.104(18) & 1.089(9) & 1.091(27) \\
\hline\hline
\end{tabular}
\label{tab:zq}
\end{center}
\end{table}
The final results in the chiral limit obtained through the two intermediate schemes are in good agreement.

The systematic uncertainties of $Z_q^{\msbar}(2\mbox{ GeV})/Z_A$ are given in
table~\ref{tab:zq_error}.
\begin{table}
\begin{center}
\caption{Systematic uncertainties of $Z_q^\msbar/Z_A(2$ GeV) in the chiral limit through the MOM and SMOM schemes for HYP1.}
\begin{tabular}{lcc}
\hline\hline
Source &  MOM (\%) &  SMOM (\%)  \\
\hline
Conversion ratio           & 0.06 & 0.12 \\
$\Lambda_{\rm QCD}^\msbar$ & 0.07 & 0.05 \\
Lattice spacing            & 0.09 & 0.01 \\
Perturbative running       &  0.10 & 0.10 \\
Fit range of $a^2p^2$      & 0.18 & 0.16 \\
Different fit ansatz       & 0.55 & 0.73 \\
$m_s^{\rm sea}\neq0$       & 0.26 & 0.20 \\
Extrapolation in $m_l$     & 0.11 & 0.10 \\
Total systematic uncertainty     & 0.66 & 0.80 \\
\hline\hline
\end{tabular}
\label{tab:zq_error}
\end{center}
\end{table}
Then we find
\begin{eqnarray}
Z_q^\msbar(2\mbox{ GeV})/Z_A&=&1.097(22)(07)\mbox{ (via MOM),}\label{eq:zq-mom}\\
Z_q^\msbar(2\mbox{ GeV})/Z_A&=&1.091(27)(09)\mbox{ (via SMOM),}\label{eq:zq-smom}
\end{eqnarray}
where the first error is statistical and the second error systematic. The dominant error is
statistical due to the $a^2p^2$-extrapolation with a second-order polynomial.
Adding up all the errors quadratically and using $Z_A=1.0789(10)$ from section~\ref{sec:zawi}, we get
\begin{equation}
Z_q^\msbar(2\mbox{ GeV})=1.184(25),
\end{equation}
where we have taken the value via the MOM scheme as our final result.
This number agrees with the $Z_q^\msbar(2\mbox{ GeV})=1.188(5)$ for ensemble 64I in Ref.~\cite{He:2022lse} (table XVII).

\section{Test the renormalization}\label{sec:twopt}

\begin{table}
\begin{center}
\caption{$Z_{A}$ and  $Z_{S,T}^{\msbar}(2\mbox{ GeV})$ on ensemble f004 for the three smearing
cases. The two errors are for the statistics uncertainty and the systematic ones.  We dropped the systematic uncertainty from the conversion ratio for $Z_S$ as it is a perturbative correction and independent of the HYP-smearing steps.}
\begin{tabular}{cccc}
\hline\hline
 & Thin & HYP1 & HYP2 \\
\hline
$Z_A$ & 1.4403(6) & 1.0808(10) & 1.0506(6) \\
$Z_S^{\msbar}(2\mbox{ GeV})$ & 1.233(11)(13) & 1.011(2)(11) & 1.004(1)(11) \\
$Z_T^{\msbar}(2\mbox{ GeV})$ & 1.576(1)(9) & 1.157(1)(6) & 1.110(1)(6) \\
\hline\hline
\end{tabular}
\label{tab:z_hyp}
\end{center}
\end{table}

In this section, we check the HYP smearing dependence of the renormalized light quark mass and hadron matrix elements on the f004 ensemble, based on the renormalization constants obtained in the previous section and collected in Table~\ref{tab:z_hyp} (without the chiral extrapolation of the sea quark mass). Besides the systematic uncertainty from the sea quark mass extrapolation, we also dropped the systematic uncertainty from the conversion ratio for $Z_S$ since it is independent of the HYP smearing setup.

\begin{figure}[htbp]
\centering
\includegraphics[width=0.49\textwidth]{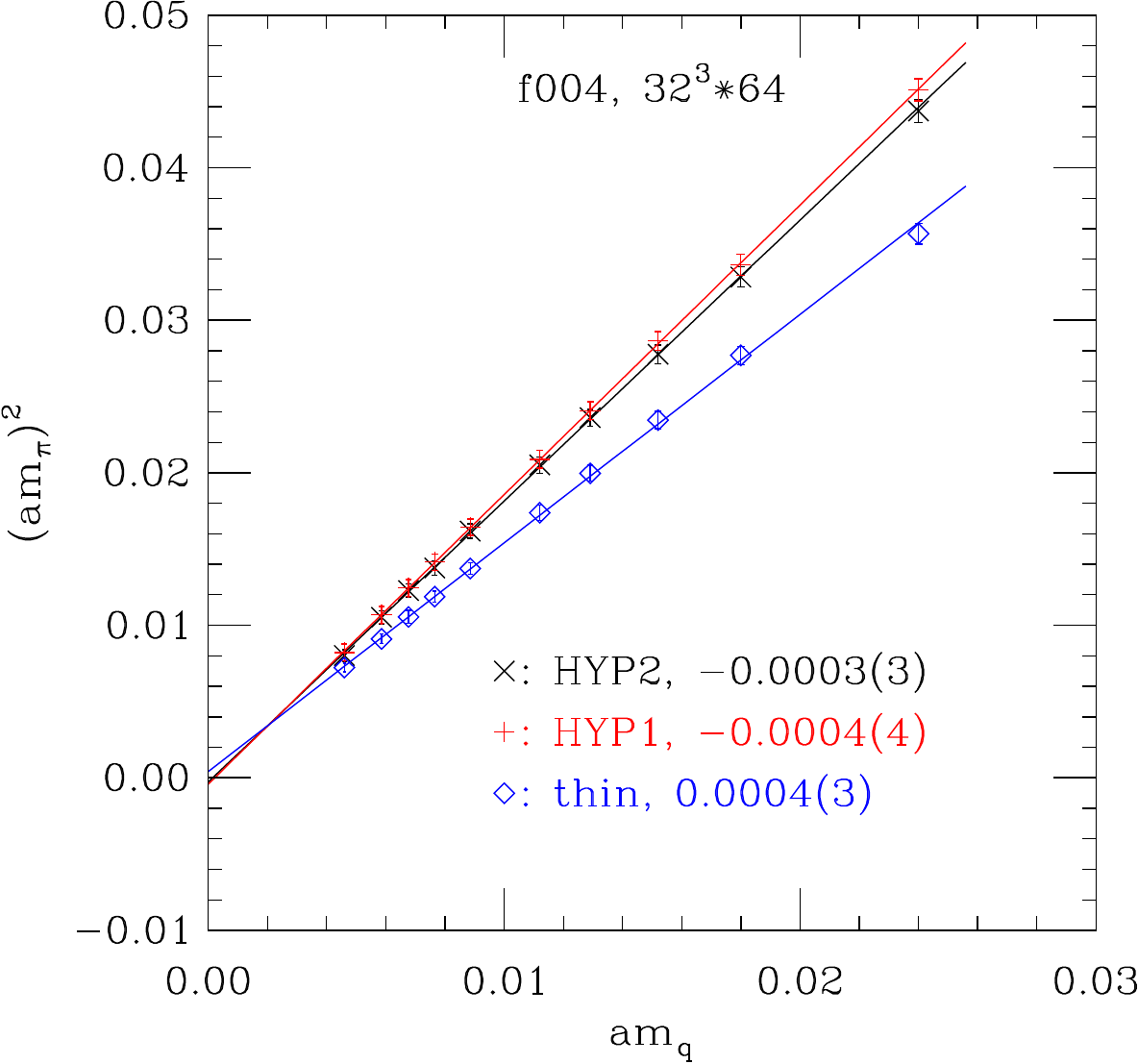}
\includegraphics[width=0.49\textwidth]{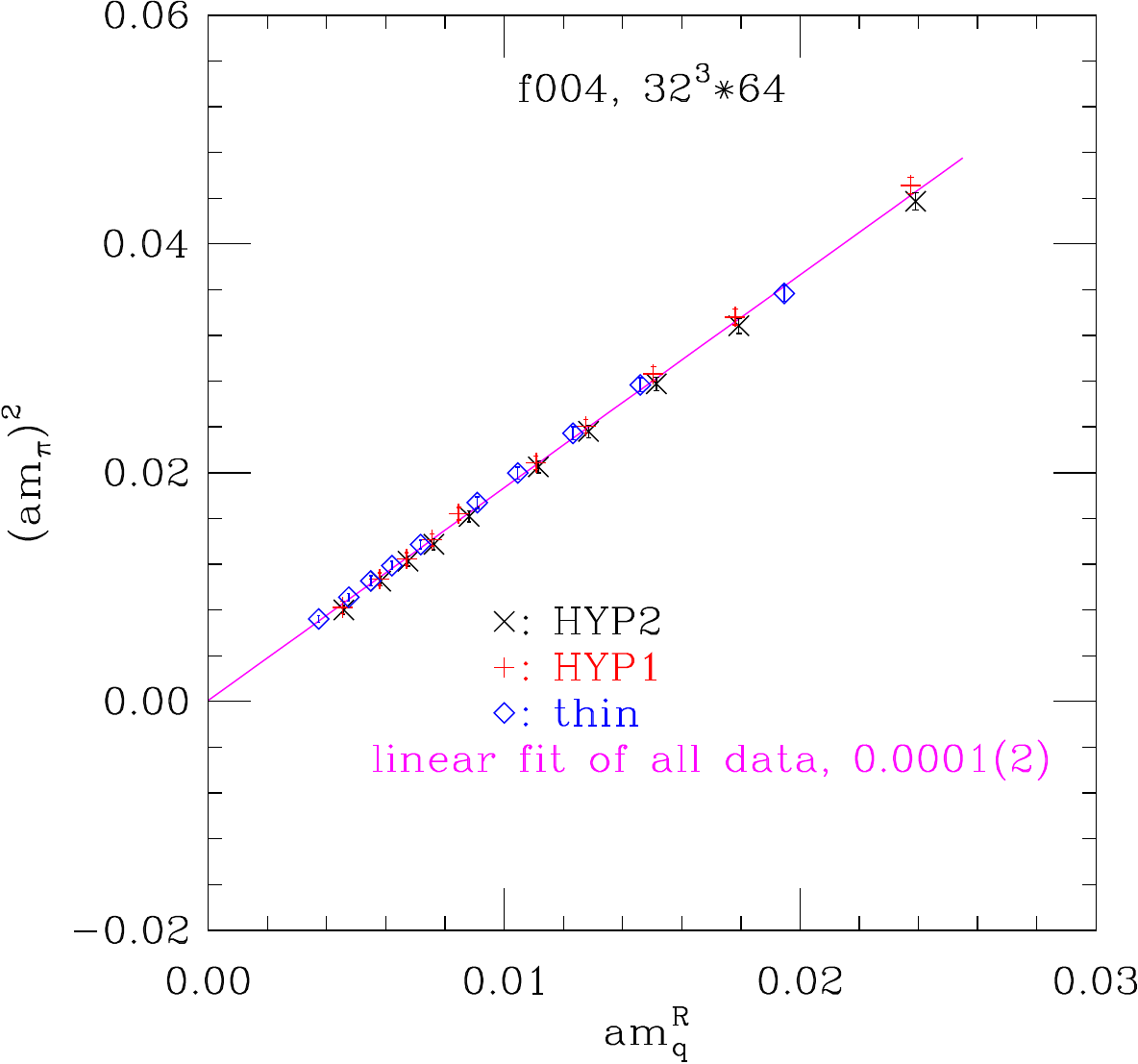}
\caption{Squared pion mass as a function of the bare quark mass (left panel) or of the renormalized quark mass $m_q^R=m_q/Z_S$ (right panel) for the three smearing cases on ensemble f004. The straight lines in the left panel are linear fits for the different smearing cases. In the right panel all data points are fitted by one straight line. The intercepts of all fits are consistent with zero as given in the graphs.}
 \label{fig:pion-mass}
\end{figure}
The quark mass dependence of the pion mass squared is shown in figure~\ref{fig:pion-mass} on the ensemble f004. About $42$ configurations are used for all three smearing cases. In the left panel we do linear fits $(am_\pi)^2=A\cdot am_q + B$ separately for the three cases, where $m_q$ is the bare quark mass. The data can be well described by this linear function and the intercepts $B$ are all consistent with zero with our statistical uncertainties. At a same bare quark mass the pion masses are different because the renormalized quark masses are different (and the discretizaton effects are also different) for the different smearing cases.

In the right panel of figure~\ref{fig:pion-mass} we plot the pion mass squared as a function of the renormalized quark mass $m_q^R(2\mbox{ GeV};\msbar)=m_q/Z_S^\msbar(2\mbox{ GeV})$, where $Z_S^\msbar(2\mbox{ GeV})$ are given in table~\ref{tab:z_hyp}. Then we find that all data points can be fitted to one linear function $(am_\pi)^2=A'\cdot am_q^R + B'$. Thus, after applying renormalization for the different smearing cases one can expect to obtain a renormalized light quark mass independent of smearing with the given statistics.

Because of the correlation between the data with different HYP smearing steps, the independence of smearing can be checked with even higher precision. We calculate the point 4-4-4 grid source~\cite{xQCD:2010pnl} propagators on two time slices on about 40 configurations, and construct the meson correlators
with different gamma matrices. Such a setup is equivalent to $4^3\times2=128$ measurements per configuration of the following point source meson correlator at short distance
\begin{align}\label{eq:c2_form}
C^R_{2,{\cal O}}(t)\equiv Z_{\cal O}^2\langle \sum_{\vec{x}} {\cal O}(\vec{x},t) {\cal O}^{\dagger}(\vec{0},0)\rangle=\sum_i \frac{\langle {\cal O}|i\rangle_R^2}{2m_i}e^{-m_it}\ _{\overrightarrow{t\rightarrow \infty}} \frac{\langle {\cal O}|0\rangle_R^2}{2m_0}e^{-m_0t},
\end{align}
where $m_i$ and $\langle {\cal O}|i\rangle$ are the $i$th state mass and matrix elements, respectively. Since the HYP smearing can change the UV behavior of the hadron spectrum, we will concentrate on the impact of HYP smearings on the ground state mass $m_0$ and also the renormalized ground state matrix element $\langle {\cal O}|0\rangle_R=Z_{\cal O}\langle {\cal O}|0\rangle$.

On a lattice with finite size $T$ in the time direction, Eq.~(\ref{eq:c2_form}) should be modified into
\begin{align}\label{eq:c2_form_lat}
C^R_{2,{\cal O}}(t)&=\sum_i \frac{\langle {\cal O}|i\rangle_R^2}{2m_i}\big[e^{-m_it}+e^{-m_i(T-t)}\big]\nonumber\\
& _{\overrightarrow{t(m_1-m_0)\gg 1}} \frac{\langle {\cal O}|0\rangle^2_R}{2m_0}\big[e^{-m_0t}+e^{-m_0(T-t)}\big].
\end{align}
For the pseudoscalar correlator which has good signal around $T/2$, we apply the one-state fit with the following ansatz in the range $0\ll t\ll T$,
\begin{align}\label{eq:c2_form_lat_1}
C^R_{2,{\cal O}=P/A4}(t)&=\frac{\langle {\cal O}|0\rangle_R^2}{2m_0}\big[e^{-m_0t}+e^{-m_0(T-t)}\big],
\end{align}
and do the two-state fit for the cases of other hadrons,
\begin{align}\label{eq:c2_form_lat_2}
C^R_{2,{\cal O}\neq P/A4}(t)&=\frac{\langle {\cal O}|0\rangle_R^2}{2m_0}\big[e^{-m_0t}(1+c_1e^{-\delta mt})+e^{-m_0(T-t)}(1+c_1e^{-\delta m(T-t)})\big],
\end{align}
at relatively smaller $t$ as the results around $T/2$ can be very noisy, where $c_1$ and $\delta m$ are additional parameters to describe the contaminations from the excited states. We do the folding $\bar{C}(t)=\frac{1}{2}[C(t)+C(T-t)]$ on the correlator and require the $\bar{C}(t)$ at the maximum $t$ used in the fit to have at least 3-5$\sigma$ signal (or $T/2$ in the pseudoscalar case), and tune the minimum $t$ to make the $\chi^2$/d.o.f. to be around one and the corresponding Q value of the fit to be larger than $0.05$.

Because of the discretization error, the meson mass using the same renormalized quark mass can be different with different HYP-smearing steps. Thus we can consider this effect in a reversed way, by tuning the quark masses to make the corresponding pseudoscalar masses obtained by $C_{2,P}$ as the following four values regardless of the HYP-smearing steps:

1) 0.302 GeV which corresponds to the unitary pion mass on the f004 ensemble;

2) 0.675 GeV which corresponds to the strange quark mass, $m_q^{\msbar}(2\mbox{ GeV})\sim 0.1\mbox{ GeV}$;

3) 0.976 GeV which corresponds to $m_q^{\msbar}(2\mbox{ GeV})\sim 0.2\mbox{ GeV}$;

4) 1.230 GeV which corresponds to $m_q^{\msbar}(2\mbox{ GeV})\sim 0.3\mbox{ GeV}$.

Practically we calculate at two quark masses around each of the above cases, and do the interpolation to make the pseudoscalar mass to be exact.

\begin{figure}[htbp]
\centering
\includegraphics[width=0.6\textwidth]{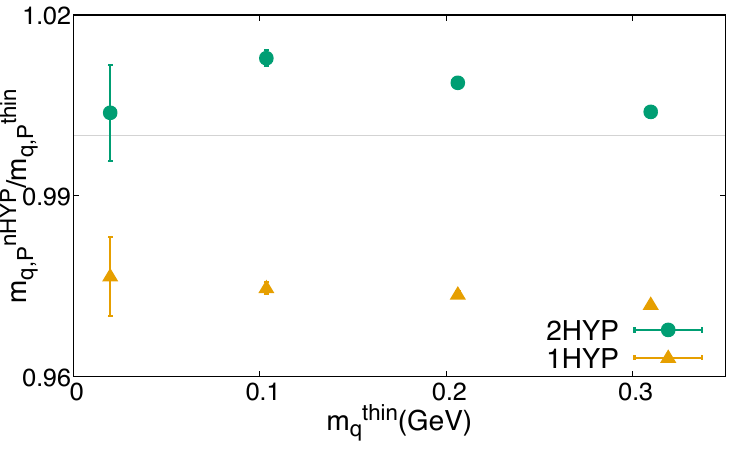}
\caption{Ratio $m_q^{n-{\rm HYP}}/m_q^{\rm thin}$ of the renormalized quark masses with $n=1$ (yellow triangles) and $n=2$ (green dots) steps of HYP smearing, to make the corresponding pseudoscalar meson mass to be 0.302, 0.675, 0.976 and 1.23 GeV, respectively.}
 \label{fig:quark_mass}
\end{figure}

Then we compare the renormalized quark mass with different HYP-smearing steps. The ratio of the quark masses with one or two steps HYP smearing over that with the thin link are plotted in  Fig.~\ref{fig:quark_mass}. The results with different HYP smearing are based on the same configurations, and the data correlation has been taken into account to suppress the uncertainty of the ratios.

In Fig.~\ref{fig:quark_mass}, the yellow triangles show the ratio of the quark mass with one-step HYP smearing (the standard $\chi$QCD setup). We can see that the ratio is around 2\%-3\% smaller than 1. Such a difference is slightly larger than the joint uncertainty of $Z_S$ with the thin link, but an order of magnitude smaller than that of the bare quark masses with or without HYP smearing ($\sim$ 20\%). On the other hand, the green dots show the case using the quark mass with two-step HYP smearing in the numerator. The deviation seems to be smaller comparing to the one-HYP case, and the result at unitary pion mass seems to be consistent with one with much larger statistical uncertainty. Thus, we conclude that the HYP smearing may introduce around 3\% discretization error in the renormalized quark mass at this specific lattice spacing.

Then we are ready to compare the renormalized ground state matrix elements $\langle {\cal O}|0\rangle$ with different ${\cal O}$, through the ratio $R_{\cal O}^{n=1,2}=\langle {\cal O}|0\rangle^{\rm nHYP}_R/\langle {\cal O}|0\rangle^{\rm thin}_R$. $R=1$ means the renormalization matrix elements are independent of the HYP smearing steps, and a nonvanishing deviation suggests a discretization error as the HYP smearing dependence should vanish in the continuum.

\begin{figure}[htbp]
\centering
\includegraphics[width=0.45\textwidth]{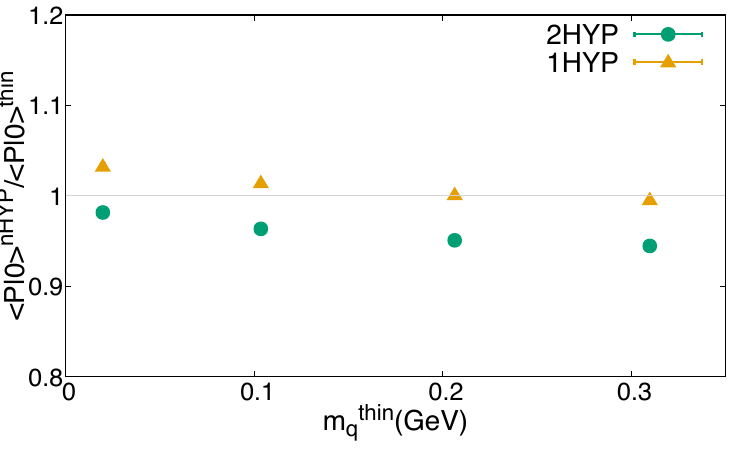}
\includegraphics[width=0.45\textwidth]{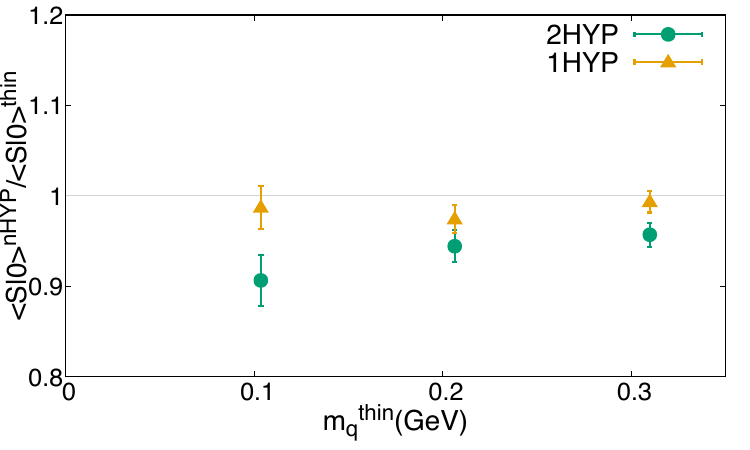}
\caption{The ratio $R_{\cal O}^{n=1,2}=\langle {\cal O}|0\rangle^{n{\rm HYP}}_R/\langle {\cal O}|0\rangle^{\rm thin}_R$ for the pseudoscalar (left panel, ${\cal O}=\bar{q}\gamma_5q$) and scalar (right panel, ${\cal O}=\bar{q}q$) matrix elements using $Z_S=Z_P$. The lightest quark mass case in the scalar channel is dropped due to the mixture between the single-hadron and multiple-hadron states.}
 \label{fig:scalar}
\end{figure}

Fig.~\ref{fig:scalar} shows the cases of the pseudoscalar (left panel) and scalar (right panel) matrix elements, while the pseudoscalar case has much better signal and shows an obvious quark mass dependence. After the chiral extrapolation, the deviation from one in the one-HYP pseudoscalar case is around 3\%, while the two-HYP case is around 1\%. The uncertainty in the scalar case is much larger and the result after chiral extrapolation is consistent with the pseudoscalar case. Note that the lightest quark mass case in the scalar channel is dropped since the mixture between the $\sigma$ and $\pi\pi$ states makes the two-state fit defined in Eq.~(\ref{eq:c2_form_lat_2}) to be unreliable. For similar reason, we dropped the light quark mass cases in all the channels except the pseudoscalar ones.

\begin{figure}[htbp]
\centering
\includegraphics[width=0.45\textwidth]{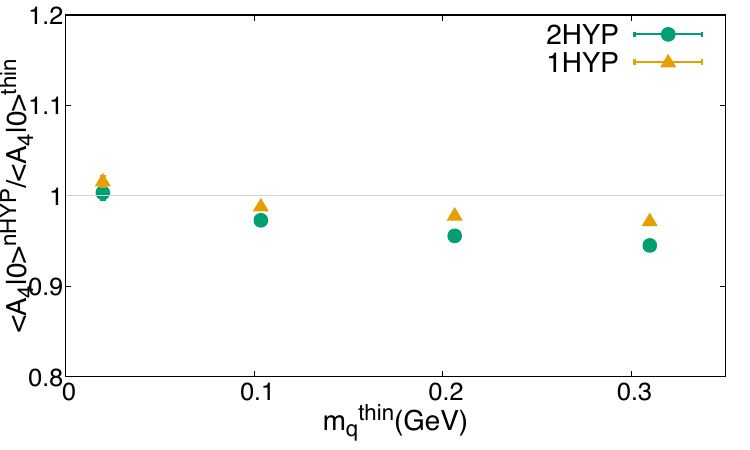}
\includegraphics[width=0.45\textwidth]{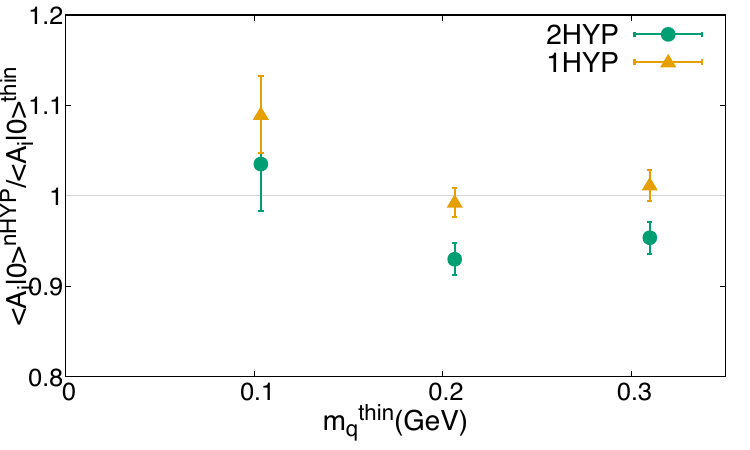}
\includegraphics[width=0.45\textwidth]{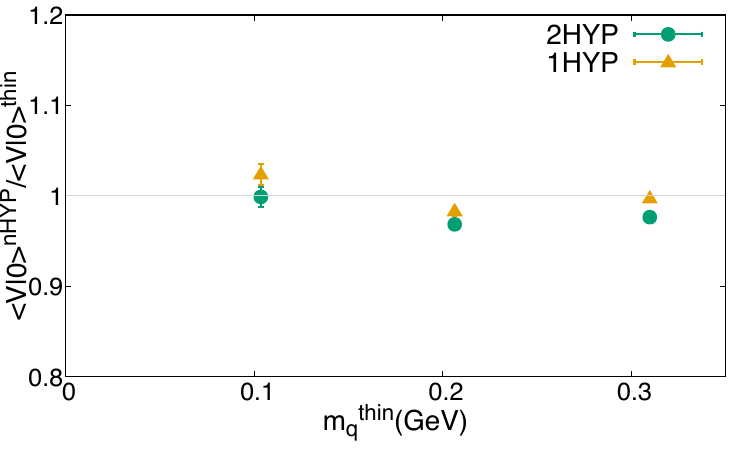}
\caption{The ratio $R_{\cal O}^{n=1,2}=\langle {\cal O}|0\rangle^{n{\rm HYP}}_R/\langle {\cal O}|0\rangle^{\rm thin}_R$ for pseudoscalar (upper left panel,  ${\cal O}=\bar{q}\gamma_5\gamma_4 q$), axial-vector (upper right panel, ${\cal O}=\bar{q}\gamma_5\gamma_iq$), and vector (lower panel, ${\cal O}=\bar{q}\gamma_iq$) matrix elements using $Z_V=Z_A$. The lightest quark mass cases in the vector and axial-vector channels are dropped due to the mixture between the single-hadron and multiple-hadron states.}
 \label{fig:vector}
\end{figure}

In the channels using the vector or axial-vector currents, the pseudoscalar channel using ${\cal O}=\bar{q}\gamma_5\gamma_4 q$ also has good signal (upper left panel of Fig.~\ref{fig:vector}) and suggests that the renormalized matrix elements in the chiral limit is independent of the HYP smearing, for both the one-HYP and two-HYP cases. Similarly to the scalar case, the results from the spatial components of the vector and axial-vector currents are noisier as shown in the other panels of Fig.~\ref{fig:vector}, while the discretization error due to the HYP smearing is at the 5\% level in the cases we investigated.

\begin{figure}[htbp]
\centering
\includegraphics[width=0.45\textwidth]{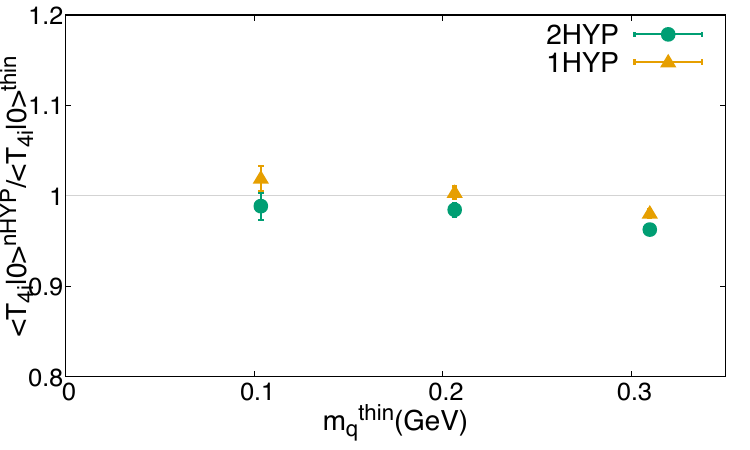}
\includegraphics[width=0.45\textwidth]{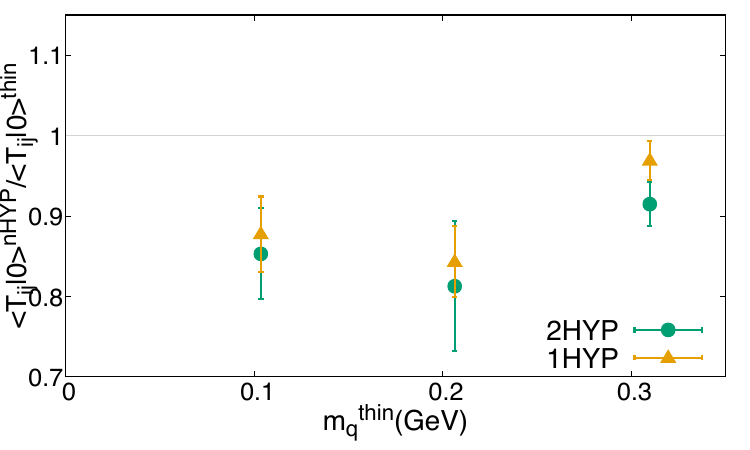}
\caption{The ratio $R_{\cal O}^{n=1,2}=\langle {\cal O}|0\rangle^{n{\rm HYP}}_R/\langle {\cal O}|0\rangle^{\rm thin}_R$ for vector (left panel, ${\cal O}=\bar{q}\gamma_4\gamma_iq$) and tensor (right panel,  ${\cal O}=\bar{q}\gamma_i\gamma_jq$) matrix elements using $Z_T$. The lightest quark mass cases are dropped due to the mixture between the single-hadron and multiple-hadron states.}
 \label{fig:tensor}
\end{figure}

As shown in Fig.~\ref{fig:tensor}, the operator ${\cal O}=\bar{q}\gamma_4\gamma_iq$ corresponds to the vector channel which is lighter than the tensor channel using the operator ${\cal O}=\bar{q}\gamma_i\gamma_jq$, and then has a better signal. We can see that the HYP smearing dependence in the vector channel is at the same level as the other cases, while the tensor channel seems to have larger deviations but the uncertainties are also large.

\section{Summary}
\label{sec:summary}
Renormalization constants are necessary in lattice QCD calculations of various hadronic matrix
elements, which are important in precise determinations of the parameters of the Standard Model and in searching new physics. Thus, one needs to calculate RCs as precise as possible.

Gauge link smearings are widely used in lattice QCD calculations. They can suppress the ultraviolet fluctuations of the gauge fields and decrease the statistical uncertainties in practice calculations
besides bringing in many other benefits. They can also affect the vertex functions of quark operators which are used in (S)MOM renormalization of those operators. Therefore we investigate the effects of HYP smearing on the renormalization window in this work as we try to get RCs
as precise as possible by using both MOM and SMOM as intermediate schemes.
We check the $a^2p^2$ dependence of $Z_X^\msbar$(2 GeV; $a^2p^2$) ($X=S, T, q$) as it is obtained
by running from the initial scale $|p|$ to 2 GeV.

For our lattice setup, in general, the $a^2p^2$ dependence of $Z_X^\msbar$(2 GeV; $a^2p^2$) via the MOM scheme is not very sensitive
to the HYP smearing although it is affected by the levels of smearing that we use (no smearing, one hit and two hits). The $a^2p^2$ dependence of $Z_{S, T}^\msbar$(2 GeV; $a^2p^2$)
can be described straightforwardly by a linear function in the range $a^2p^2\in [5, 10]$
for all smearing cases. In physical units the range is $|p|\in [5.3, 7.5]$ GeV.
The $a^2p^2$ dependence of $Z_{q}^\msbar$(2 GeV; $a^2p^2$) can be described by a second-order polynomial for all three smearing cases.

For the intermediate SMOM scheme, the behaviour of the $a^2p^2$ dependence of $Z_X^\msbar$(2 GeV; $a^2p^2$)
is sensitive to whether HYP smearing is applied or not. From the right panels of
figures~\ref{fig:zs-smom}, \ref{fig:zt-smom} and~\ref{fig:zq} we can see
that the behaviors of $Z_X^\msbar$(2 GeV; $a^2p^2$) for HYP1 and HYP2 are apparently
different from that for the thin case. $Z_X^\msbar$(2 GeV; $a^2p^2$) can be described by a
linear function of $a^2p^2$ when no smearing is applied. Among the three RCs ($X=S, T, q$)
the $a^2p^2$ dependence of $Z_{T}^\msbar$(2 GeV; $a^2p^2$) has the least sensitivity to
smearing. It can still be fitted to linear functions for both HYP1 and HYP2 although the slope
has a sizable change compared with no smearing.
The $a^2p^2$ dependence of $Z_{q}^\msbar$(2 GeV; $a^2p^2$)
changes from a linear function to a second-order polynomial after smearing is applied.
$Z_{S}^\msbar$(2 GeV; $a^2p^2$) has the largest sensitivity to smearing. With HYP2 we find it is hard to describe the $a^2p^2$ dependence of $Z_{S}^\msbar$(2 GeV; $a^2p^2$) no matter if we
use high order terms for discretization effects or inverse terms for possible nonperturbative effects.

By going to higher renormalization scale than in~\cite{Liu:2013yxz} and using four-loop
conversion ratios from the MOM scheme to $\msbar$, we reduce the
systematic errors of the RCs for the scalar and tensor operators.
For $Z_{T}^\msbar(2\mbox{ GeV})$ we obtain a total uncertainty less than $1\%$.

We also checked the HYP smearing dependence of the renormalized quark masses and hadron matrix elements. The results show that the renormalization suppresses the $\sim$ 30\% difference of the bare quantities into 3\%-5\% level, while more than one step of HYP smearing may not make the residual deviation to be smaller.

\section*{Acknowledgements}
We thank RBC-UKQCD collaborations for sharing the domain wall fermion configurations.
This work is supported in part by National Key Research and Development Program of China under Contract No. 2020YFA0406400 and by the National Natural Science Foundation of China (NNSFC) under Grants No. 12075253, 11935017, 12192264, 12293060, 12293065, 12047503, 12293062 and 12070131001 (CRC 110 by DFG and NNSFC). K. L. is supported by the U.S. DOE Grant No. DE-SC0013065 and DOE Grant No. DEAC05-06OR23177, which is within the framework of the TMD Topical Collaboration.
Y. Y. is supported in part by the Strategic Priority Research Program of Chinese Academy of Sciences under Grants No.\ XDB34030303 and XDPB15, and a NSFC-DFG joint grant under Grant No. 12061131006 and SCHA 458/22. The GWU code~\cite{Alexandru:2011ee,Alexandru:2011sc} is acknowledged. The computations were performed on the HPC clusters at Institute of High Energy Physics (Beijing) and China Spallation Neutron Source (Dongguan).
Part of the calculations were performed through HIP programming model~\cite{Bi:2020wpt} on the ORISE Supercomputer.


\begin{thebibliography}{99}
\bibitem{Chen:2020qma}
Y.~Chen \textit{et al.} [\ensuremath{\chi}QCD],
``Charmed and $\phi$ meson decay constants from 2+1-flavor lattice QCD,''
Chin. Phys. C \textbf{45}, no.2, 023109 (2021)
\href{https://doi.org/10.1088/1674-1137/abcd8f}{doi:10.1088/1674-1137/abcd8f}
[arXiv:2008.05208 [hep-lat]].

\bibitem{Hatton:2020vzp}
D.~Hatton \textit{et al.} [HPQCD],
``Renormalization of the tensor current in lattice QCD and the $J/\psi$ tensor decay constant,''
Phys. Rev. D \textbf{102}, no.9, 094509 (2020)
doi:10.1103/PhysRevD.102.094509
[arXiv:2008.02024 [hep-lat]].

\bibitem{Aoki:2007xm}
  Y.~Aoki, P.~A.~Boyle, N.~H.~Christ, C.~Dawson, M.~A.~Donnellan, T.~Izubuchi, A.~Juttner and S.~Li {\it et al.},
  Phys.\ Rev.\ D {\bf 78}, 054510 (2008)
  \href{http://arxiv.org/abs/0712.1061}{\tt [arXiv:0712.1061 [hep-lat]]}.

\bibitem{Sturm:2009kb}
  C.~Sturm, Y.~Aoki, N.~H.~Christ, T.~Izubuchi, C.~T.~C.~Sachrajda and A.~Soni,
  Phys.\ Rev.\ D {\bf 80}, 014501 (2009)
  \href{http://arXiv.org/abs/0901.2599}{\tt [arXiv:0901.2599 [hep-ph]]}.

\bibitem{Martinelli:1994ty}
  G.~Martinelli, C.~Pittori, C.~T.~Sachrajda, M.~Testa and A.~Vladikas,
  Nucl.\ Phys.\  B {\bf 445} (1995) 81
  \href{http://arXiv.org/abs/hep-lat/9411010}{\tt [arXiv:hep-lat/9411010]}.

\bibitem{Gorbahn:2010bf}
  M.~Gorbahn and S.~J\"ager,
  Phys.\ Rev.\ D {\bf 82}, 114001 (2010)
  \href{http://arXiv.org/abs/1004.3997}{\tt [arXiv:1004.3997 [hep-ph]]}.

\bibitem{Almeida:2010ns}
  L.~G.~Almeida and C.~Sturm,
  Phys.\ Rev.\ D {\bf 82}, 054017 (2010)
  \href{http://arXiv.org/abs/1004.4613}{\tt [arXiv:1004.4613 [hep-ph]]}.

\bibitem{Kniehl:2020sgo}
B.~A.~Kniehl and O.~L.~Veretin,
Phys. Lett. B \textbf{804}, 135398 (2020)
doi:10.1016/j.physletb.2020.135398
[arXiv:2002.10894 [hep-ph]].

\bibitem{Bednyakov:2020ugu}
A.~Bednyakov and A.~Pikelner,
``Quark masses: N3LO bridge from ${\rm RI/SMOM}$ to ${\rm \overline{MS}}$ scheme,''
Phys. Rev. D \textbf{101}, no.9, 091501 (2020)
doi:10.1103/PhysRevD.101.091501
[arXiv:2002.12758 [hep-ph]].

\bibitem{Bi:2017ybi}
Y.~Bi, H.~Cai, Y.~Chen, M.~Gong, K.~F.~Liu, Z.~Liu and Y.~B.~Yang,
Phys. Rev. D \textbf{97}, no.9, 094501 (2018)
doi:10.1103/PhysRevD.97.094501
[arXiv:1710.08678 [hep-lat]].

\bibitem{He:2022lse}
F.~He \textit{et al.} [\ensuremath{\chi}QCD],
Phys. Rev. D \textbf{106}, no.11, 114506 (2022)
doi:10.1103/PhysRevD.106.114506
[arXiv:2204.09246 [hep-lat]].

\bibitem{Hasenfratz:2001hp}
A.~Hasenfratz and F.~Knechtli,
Phys. Rev. D \textbf{64}, 034504 (2001)
doi:10.1103/PhysRevD.64.034504
[arXiv:hep-lat/0103029 [hep-lat]].

\bibitem{xQCD:2010pnl}
A.~Li \textit{et al.} [xQCD],
``Overlap Valence on 2+1 Flavor Domain Wall Fermion Configurations with Deflation and Low-mode Substitution,''
Phys. Rev. D \textbf{82}, 114501 (2010)
doi:10.1103/PhysRevD.82.114501
[arXiv:1005.5424 [hep-lat]].

\bibitem{Hasenfratz:2007rf}
A.~Hasenfratz, R.~Hoffmann and S.~Schaefer,
``Hypercubic smeared links for dynamical fermions,''
JHEP \textbf{05}, 029 (2007)
doi:10.1088/1126-6708/2007/05/029
[arXiv:hep-lat/0702028 [hep-lat]].

\bibitem{Arthur:2013bqa}
R.~Arthur, P.~A.~Boyle, S.~Hashimoto and R.~Hudspith,
``Note on Rome-Southampton renormalization with smeared gauge fields,''
Phys. Rev. D \textbf{88}, no.11, 114506 (2013)
doi:10.1103/PhysRevD.88.114506
[arXiv:1306.0835 [hep-lat]].

\bibitem{RBC:2010qam}
Y.~Aoki, R.~Arthur \textit{et al.} [RBC and UKQCD],
Phys. Rev. D \textbf{83}, 074508 (2011)
doi:10.1103/PhysRevD.83.074508
[arXiv:1011.0892 [hep-lat]].

\bibitem{Liu:2013yxz}
Z.~Liu \textit{et al.} [chiQCD],
``Nonperturbative renormalization of overlap quark bilinears on 2+1-flavor domain wall fermion configurations,''
Phys. Rev. D \textbf{90}, no.3, 034505 (2014)
doi:10.1103/PhysRevD.90.034505
[arXiv:1312.7628 [hep-lat]].

\bibitem{Aoki:2010dy}
  Y.~Aoki {\it et al.}  [RBC and UKQCD Collaborations],
  Phys.\ Rev.\ D {\bf 83}, 074508 (2011)
  \href{http://arxiv.org/abs/1011.0892}{\tt [arXiv:1011.0892 [hep-lat]]}.

\bibitem{Neuberger:1997fp}
  H.~Neuberger,
  Phys.\ Lett.\ B {\bf 417}, 141 (1998)
  \href{http://arxiv.org/abs/hep-lat/9707022}{\tt [hep-lat/9707022]}.

\bibitem{Chiu:1998gp}
  T.~-W.~Chiu and S.~V.~Zenkin,
  Phys.\ Rev.\ D {\bf 59}, 074501 (1999)
  \href{http://arxiv.org/abs/hep-lat/9806019}{\tt [hep-lat/9806019]}.

\bibitem{Franco:1998bm}
  E.~Franco and V.~Lubicz,
  Nucl.\ Phys.\  B {\bf 531} (1998) 641
  \href{http://arXiv.org/abs/hep-ph/9803491}{\tt [arXiv:hep-ph/9803491]}.

\bibitem{Chetyrkin:1999pq}
  K.~G.~Chetyrkin and A.~Retey,
  ``Renormalization and running of quark mass and field in the  regularization
  Nucl.\ Phys.\  B {\bf 583} (2000) 3
  \href{http://arxiv.org/abs/hep-ph/9910332}{\tt [arXiv:hep-ph/9910332]}.

\bibitem{Gracey:2022vjc}
J.~A.~Gracey,
[arXiv:2210.12420 [hep-ph]].

\bibitem{RBC:2014ntl}
T.~Blum \textit{et al.} [RBC and UKQCD],
Phys. Rev. D \textbf{93}, no.7, 074505 (2016)
doi:10.1103/PhysRevD.93.074505
[arXiv:1411.7017 [hep-lat]].

\bibitem{ParticleDataGroup:2016lqr}
C.~Patrignani \textit{et al.} [Particle Data Group],
Chin. Phys. C \textbf{40}, no.10, 100001 (2016)
doi:10.1088/1674-1137/40/10/100001

\bibitem{Alekseev:2002zn}
  A.~I.~Alekseev,
  Few Body Syst.\  {\bf 32}, 193 (2003)
  \href{http://arxiv.org/abs/hep-ph/0211339}{\tt [hep-ph/0211339]}.

\bibitem{vanRitbergen:1997va}
  T.~van Ritbergen, J.~A.~M.~Vermaseren and S.~A.~Larin,
  Phys.\ Lett.\ B {\bf 400}, 379 (1997)
  \href{http://arxiv.org/abs/hep-ph/9701390}{\tt [hep-ph/9701390]}.

\bibitem{Horkel:2020hpi}
D.~Horkel \textit{et al.} [\ensuremath{\chi}QCD],
Phys. Rev. D \textbf{101}, no.9, 094501 (2020)
doi:10.1103/PhysRevD.101.094501
[arXiv:2002.06699 [hep-lat]].

\bibitem{Gracey:2003yr}
J.~A.~Gracey,
Nucl. Phys. B \textbf{662}, 247-278 (2003)
doi:10.1016/S0550-3213(03)00335-3
[arXiv:hep-ph/0304113 [hep-ph]].

\bibitem{Gracey:2022vqr}
J.~A.~Gracey,
``Tensor current renormalization in the RI' scheme at four loops,''
Phys. Rev. D \textbf{106}, no.8, 085008 (2022)
doi:10.1103/PhysRevD.106.085008
[arXiv:2208.14527 [hep-ph]].

\bibitem{Baikov:2006ai}
P.~A.~Baikov and K.~G.~Chetyrkin,
Nucl. Phys. B Proc. Suppl. \textbf{160}, 76-79 (2006)
doi:10.1016/j.nuclphysbps.2006.09.031

\bibitem{Sternbeck:2016ltn}
A.~Sternbeck,
EPJ Web Conf. \textbf{137}, 01020 (2017)
\href{https://doi.org/10.1051/epjconf/201713701020}{doi:10.1051/epjconf/201713701020}
[arXiv:1612.06106 [hep-lat]].

\bibitem{Blossier:2010vt}
B.~Blossier, P.~Boucaud, M.~Brinet, F.~De Soto, Z.~Liu, V.~Morenas, O.~Pene, K.~Petrov and J.~Rodriguez-Quintero,
``Renormalisation of quark propagators from twisted-mass lattice QCD at $N_f$=2,''
Phys. Rev. D \textbf{83}, 074506 (2011)
doi:10.1103/PhysRevD.83.074506
[arXiv:1011.2414 [hep-ph]].

\bibitem{Virgili:2022wfx}
A.~Virgili, W.~Kamleh and D.~Leinweber,
[arXiv:2209.14864 [hep-lat]].

\bibitem{Wang:2016lsv}
C.~Wang, Y.~Bi, H.~Cai, Y.~Chen, M.~Gong and Z.~Liu,
``Quark chiral condensate from the overlap quark propagator,''
Chin. Phys. C \textbf{41}, no.5, 053102 (2017)
doi:10.1088/1674-1137/41/5/053102
[arXiv:1612.04579 [hep-lat]].

\bibitem{Chetyrkin:1997dh}
  K.~G.~Chetyrkin,
  Phys.\ Lett.\ B {\bf 404}, 161 (1997)
  doi:10.1016/S0370-2693(97)00535-2
  \href{http://arXiv.org/abs/hep-ph/9703278}{\tt [hep-ph/9703278]}.

\bibitem{Alexandru:2011ee}
A.~Alexandru, C.~Pelissier, B.~Gamari and F.~Lee,
J. Comput. Phys. \textbf{231}, 1866-1878 (2012)
doi:10.1016/j.jcp.2011.11.003
[arXiv:1103.5103 [hep-lat]].

\bibitem{Alexandru:2011sc}
A.~Alexandru, M.~Lujan, C.~Pelissier, B.~Gamari and F.~X.~Lee,
doi:10.1109/SAAHPC.2011.13
[arXiv:1106.4964 [hep-lat]].

\bibitem{Bi:2020wpt}
Y.~J.~Bi, Y.~Xiao, W.~Y.~Guo, M.~Gong, P.~Sun, S.~Xu and Y.~B.~Yang,
PoS \textbf{LATTICE2019}, 286 (2020)
doi:10.22323/1.363.0286
[arXiv:2001.05706 [hep-lat]].

\end{thebibliography}
\end{document}